\documentclass[twocolumn]{aastex7}
\usepackage{placeins}

\usepackage{lipsum}
\usepackage{tabularx,booktabs}
\usepackage[normalem]{ulem}

\begin{document}

\title{The Critical Mass in Galaxy Evolution}

\author[0009-0009-3772-3134]{Preetish K. Mishra}
\affiliation{School of Physics, Korea Institute for Advanced Study,
85 Hoegiro, Dongdaemun-gu, Seoul 02455,
Republic of Korea}
\email[show]{preetish21@kias.re.kr}

\author[0000-0001-9521-6397]{Changbom Park}
\affiliation{School of Physics, Korea Institute for Advanced Study,
85 Hoegiro, Dongdaemun-gu, Seoul 02455,
Republic of Korea}
\email{cbp@kias.re.kr}

\author[0000-0002-6810-1778]{Jaehyun Lee}
\affiliation{Korea Astronomy and Space Science Institute, 776, Daedeokdae-ro, Yuseong-gu, Daejeon 34055, Republic of Korea}
\email[show]{jaehyun@kasi.re.kr}

\author[0000-0003-0225-6387]{Yohan Dubois}
\affiliation{Institut d’Astrophysique de Paris, UMR 7095, CNRS, Sorbonne Université, 98 bis boulevard Arago, F-75014 Paris, France}
\email{dubois@iap.fr}

\author[0000-0003-0695-6735]{Christophe Pichon}
\affiliation{Institut d’Astrophysique de Paris, UMR 7095, CNRS, Sorbonne Université, 98 bis boulevard Arago, F-75014 Paris, France}
\affiliation{Department of Astronomy \& Space Science, Kyung Hee University, 1732 Deogyeong-daero, Yongin-si, Gyeonggi-do 17104, Republic of Korea}
\email{pichon@iap.fr}

\author[0000-0002-4391-2275]{Juhan Kim}
\affiliation{Center for Advanced Computation, Korea Institute for Advanced Study, 85 Hoegiro, Dongdaemun-gu, Seoul 02455,
Republic of Korea}
\email{kjhan@kias.re.kr}

\author[0000-0003-4446-3130]{Brad Gibson}
\affiliation{Woodmansey Primary School, Hull Road, Woodmansey HU17 0TH, UK}
\email{brad.k.gibson@gmail.com}



\begin{abstract}

We investigate the physical origin of critical mass, a threshold where many galaxy properties and scaling relations undergo fundamental transitions, using the Horizon Run 5 simulation. Focusing on massive ($M_{\rm tot} \geq 10^{12}{\rm M_\odot}$) central galaxies, we examine the mass-dependent turnover of the stellar-to-total mass ratio (STR) and the physical processes driving it. We decompose STR into the stellar-to-baryon mass ratio ($M_*/M_{\rm bar}$) and baryon retention fraction ($M_{\rm bar}/M_{\rm tot}$) to examine galaxies' ability to retain baryons and convert them into stars. We find that STR evolution is dominated by variation in $M_*/M_{\rm bar}$, which changes by over a factor of three, peaking within a narrow range of $M_{\rm tot} \sim 10^{12.4\text{--}12.7}{\rm M_\odot}$ independent of redshift, while $M_{\rm bar}/M_{\rm tot}$ varies by at most 30\%. A redshift-independent critical mass at $M_{\rm tot} \sim 10^{12.5}{\rm M_\odot}$ ($M_* \sim 10^{10.7}{\rm M_\odot}$) arises from the changing nature of gas accretion. At this scale, a dynamically stable hot gas halo develops that suppresses cool gas inflow, reducing in-situ star formation efficiency such that $M_{\rm tot}$ growth exceeds in-situ $M_{*}$ growth. Consequently, the hot gas reservoir grows while $M_{*}$ growth slows, producing upturns in $M_{\rm gas}/M_{\rm tot}$ and $M_{\rm bar}/M_{\rm tot}$ and a downturn in $M_{*}/M_{\rm bar}$ that ultimately drives the STR turnover. We also identify a secondary critical mass at $M_{\rm tot} \approx 10^{11}{\rm M_\odot}$ (or $M_{*} \approx 10^{9\text{--}9.5}{\rm M_\odot}$) where gas retention fraction peaks, above which increasing hot gas fraction gradually suppresses in-situ star formation efficiency.

\end{abstract}



\section{Introduction} \label{sec:intro}

Galaxies, composed of stars, gas, and dark matter, are the fundamental building blocks of the universe. They are characterized by a wide range of physical properties, including mass, size, morphology, kinematics, color, star formation activity, and the metallicity of their stellar and gaseous components. Measuring these properties and analyzing population-averaged statistical trends provide key insights into the processes driving galaxy formation and evolution. In the era of large-scale surveys, numerous studies have investigated the interrelations among these physical properties of low-redshift galaxies \citep[see e.g.,][]{Shen2003, Kauffman2004, Baldry2004, Tremonti2004, Choi2007, Park2007, Park2008}. A central result from these studies is that most key galaxy properties are strongly correlated with the total stellar content of galaxies, commonly quantified by luminosity or stellar mass.

Among these general trends, one particularly intriguing result stands out: there exists a critical mass scale, roughly at a stellar mass of $10^{10.5} {\rm M_{\odot}}$, where the average physical properties of galaxies undergo significant transitions. Around this mass, the galaxy population shifts from predominantly star-forming (blue) to passive (red) \citep[see e.g.,][]{Wetzel2013, Davies2019}, while the transition from disk-dominated to spheroid-dominated morphology occurs over a similar stellar mass range \citep[see e.g.,][]{wel2008, Kelvin2014, Park_2022}. Several well-known scaling relations, such as the galaxy size–stellar mass relation \citep[see e.g.,][]{Mowla_2019, Nancy2021, Mishra2023}, the stellar mass–gas metallicity relation \citep[][]{Tremonti2004, Lopez2013}, stellar mass-black hole mass relation \citep[see eg.][]{Reines2015},  and the stellar mass–halo mass relation \citep[see e.g.,][]{Wechsler_2018}—exhibit a change in slope around $M_* \sim 10^{10.2}$–$10^{10.6} {\rm M_{\odot}}$. 
\cite{Choi2007} showed that the red sequence of SDSS galaxies shows a slope change at ${\rm M}_r \approx -19.6 + 5 \log h$, corresponding to stellar mass of $10^{10.5} M_{\odot}$. In particular, this luminosity also marks a change in the slope of the Faber–Jackson relation and a decrease in the intrinsic scatter of galaxy color, size, and concentration for early-type galaxies \citep{Choi2007}. Recent observations further indicate a change in the slope of the Tully–Fisher relation for late-type galaxies at a comparable luminosity \citep[see e.g.,][]{Boubel2024}. The stellar mass scale of $10^{10.5} {\rm M_{\odot}}$ also appears to be important in shaping galaxy rotation curves. \citet{Yoon_2021} found that late-type galaxies above $M_* \sim 10^{10.5} {\rm M_{\odot}}$ exhibit nearly flat outer rotation curves, while less massive systems show rising curves. For early-type galaxies, the inner rotation-curve slope peaks near this mass and decreases toward both higher and lower masses, consistent with their central stellar surface density also peaking at this scale.

Cosmological simulations reproduce the observed critical mass scale associated with integrated galaxy properties (e.g., color, star formation, and metallicity), and recent state-of-the-art hydrodynamical simulations extend this result to resolved galaxy properties. Using the Horizon Run 5 \citep{Lee2021} simulation, \citet{Park_2022} showed that disk galaxies dominate at $M_* = 10^{9}\sim 10^{10} {\rm M_{\odot}}$, while spheroid-dominated systems become prevalent above $M_* \gtrsim 10^{11} {\rm M_{\odot}}$, with a sharp rise in the spheroid fraction between $10^{10.5}$ and $10^{11} {\rm M_{\odot}}$. These trends have been independently confirmed by later JWST observations \citep{JHLee2024}. \citet{Jeong2025} further showed that the dependence of galaxy rotation-curve shapes on stellar mass and morphology can be successfully reproduced in cosmological simulations.

The existence of a common mass threshold across diverse galaxy properties suggests that galaxies above this scale encounter fundamentally different evolutionary pathways, raising the question of what physical mechanisms set the critical mass at $M_* \sim 10^{10.5} {\rm M_{\odot}}$.

Theoretical explanations have so far been limited. \citet{Dekel_2006} proposed that this characteristic stellar mass arises from a transition in gas accretion physics at a halo mass of $\sim10^{12}{\rm M_{\odot}}$, above which virial shocks heat infalling gas and suppress star formation. At $z \gtrsim 2$, cold streams can penetrate hot halos and sustain star formation, but below $z \lesssim 2$ such streams disappear, leading to quenching in massive halos. Extending this framework, \citet{Dekel_2019} linked the transition to angular-momentum loss in discs, forming compact “blue nuggets” that undergo inside-out quenching due to rapid black-hole growth and evolve into red nuggets \citep{Lapiner2021, Lapiner2023}.

A key prediction of \citet{Dekel_2019} is a nearly constant critical halo mass of $\sim10^{12} {\rm M_{\odot}}$ over $0 < z < 2$, rising rapidly at higher redshift. Notably, this framework emphasizes $z \sim 2$ as a critical epoch for quenching, likely influenced by earlier UV-based studies that identified a peak in the cosmic star formation rate density at $z \sim 2$ \citep[e.g.,][]{Madau1998, Madau_Dickinson2014}. However, longer-wavelength FIR, radio, and sub-mm observations, being less affected by dust, indicate a broader peak with an extended turnover at $2 < z < 3$  \citep[e.g.,][]{Grupponi2013, Novac2017, Gruppioni2020, Enia_2022}, suggesting that reduced star formation efficiency may begin earlier than $z \sim 2$. Similarly, while HST-based studies implied a high abundance of blue nuggets at $z > 2$, recent JWST results show that such systems are rare or constitute only a minor fraction of massive galaxies \citep[][]{Bail2024, Magnelli2023}, challenging their role as a dominant transformation phase.

Using a suite of small-volume cosmological simulations, \citet{Tortora2025} showed that the critical mass associated with the galaxy size–mass and stellar mass–halo mass relations may vary with AGN and supernova feedback parameters, although the effect is modest. However, examining how the critical mass changes with feedback alone does not reveal the detailed physical mechanisms responsible for its emergence. A more informative approach is to trace how key processes, such as gas accretion and star formation, evolve as galaxies grow toward their critical mass along their evolutionary histories. Moreover, the physical drivers of the critical mass should be examined separately for distinct galaxy properties, such as stellar mass, morphology, and kinematics, since they may arise from different processes. For example, mechanisms that regulate stellar mass growth may not be the same as those driving morphological transformation, which is also known to be strongly influenced by environment \citep[][]{Park2007, Park2009, Hong2024}.


In this work, we use the Horizon Run 5 (HR5) cosmological hydrodynamical simulation \citep{Lee2021} to investigate the origin of the critical mass scale, focusing on the stellar mass–total mass relation which is one of the most fundamental links between different components of galaxies \citep[e.g.,][]{ Behroozi2013, Kravtsov2018, Shuntov2025, Paquereau2025}. By tracing the emergence of the peak in the stellar-to-total mass ratio at $M_{\rm total} \sim 10^{12}{\rm M_{\odot}}$ (or $M_* \sim 10^{10.3}{\rm M_{\odot}}$), we aim to identify the physical processes responsible for setting this critical mass and its connection to broader galaxy evolution.

The paper is organized as follows: Section \ref{sec:data} details the galaxy sample, the measurements of key quantities, and the methodology used in our analysis. In Section \ref{sec:results}, we first examine how the stellar-to-total mass ratio, the stellar-to-baryon mass ratio and baryon retention fraction, and the total gas and ISM mass fractions evolve as galaxies grow in total mass. We then investigate the nature of gas accretion and its impact on star formation efficiency, and conclude by examining the redshift evolution of the critical mass scale. Finally, Section \ref{sec:summary} summarizes our findings.

\section{Data \& Methodology} \label{sec:data}

\subsection{The Horizon Run 5 Simulation} 

The Horizon Run 5 (HR5) simulation \citep{Lee2021} is a state-of-the-art, large-volume cosmological hydrodynamical simulation. It adopts a flat $\Lambda$CDM cosmology with parameters consistent with \textit{Planck} 2015 data \citep{Planck2015}: $\Omega_m = 0.3$, $\Omega_\Lambda = 0.7$, $\Omega_b = 0.047$, $\sigma_8 = 0.816$, and $h_0 = 0.684$. The linear power spectrum for the initial conditions was calculated using the \texttt{CAMB} package \citep{Lewis2000}, and the initial conditions were generated at a redshift of $z = 200$ using the \texttt{MUSIC} package \citep{Hahn2011}. HR5 was performed using a version of an adaptive mesh refinement code RAMSES \citep{Teyssier2002} optimized for the MPI-OpenMP hybrid parallelism \citep{Lee2021}.
It covers a co-moving volume of 1.05\,cGpc$^3$ and traces the formation and evolution of galaxies and large-scale cosmic structures up to a redshift of $z = 0.625$, achieving a spatial resolution of about 1 proper kpc within a zoomed-in cuboidal region of $1049 \times 119 \times 127$\,cMpc$^3$.

The HR5 simulation implements subgrid physics to model various physical processes including gas cooling, reionization, star formation, chemical evolution, formation and growth of massive blackholes, and AGN and supernova feedback. Further details of the implementation can be found in \cite{Lee2021}. The subgrid parameters are chosen so that the simulated cosmic star formation history matches the observational data collected from \cite{Hopkins2004}, \cite{Behroozi2013}, and \cite{Madau_Dickinson2014}.

\subsection{Galaxy Catalogue \& Data}
The HR5 simulation employs an extensive approach to identify galaxies and dark matter haloes. First, an extended FoF algorithm, based on an adoptive linking length scheme, was applied to the combined set of dark matter, stars, BHs, and gas grids to identify FoF halos. Then, a PSB \citep{Kim2006}-based galaxy finder, PGalF \citep[see for details][]{Kim2023}, was applied to find substructures within the FoF haloes. The PGalF searches for local density peaks of stellar and dark matter within the FoF halos to identify self-bound substructures. A substructure is defined as a galaxy when it contains sufficient stellar particles for constructing a stellar-mass density field, and then a galaxy catalogue is created for each snapshot \citep[see for details][]{Lee2021}.

The galaxy catalogue in the HR5 database contains a range of measurements of diverse properties of galaxies, as well as their group membership information. In this work, we mainly employ the masses of baryonic components bound to substructures. These are: stellar mass ($M_{*}$), gas mass ($M_{\rm gas}$), dark matter mass ($M_{\rm DM}$), and mass of the black hole ($M_{\rm BH}$). Based on these quantities, we define total mass of the substructure as $M_{\rm tot} = M_{*} + M_{\rm gas} + M_{\rm DM} + M_{\rm BH}$. From this point onward, we will simply refer $M_{\rm tot}$ as the total mass of galaxies\footnote{While often called “halo mass” in the literature, we avoid this term because our definition includes the total bound mass of all components—dark matter, stars, gas, and black holes—across the halo, disk, bulge, and nucleus. Even common mass definitions like $M_{200c}$ include more than just dark matter. The term “halo mass” is a legacy of the early $N$-body simulation era and should be avoided unless one specifically intends to refer to one of the dark matter, stellar, or gas halo by `halo'.}.

In addition to substructure-wide quantities, we define the interstellar gas mass, $M_{\rm ISM}$, as the gas mass enclosed within five times the stellar half-mass radius ($5R_{1/2}$), which approximately corresponds to the main stellar extent of galaxies. This completes the list of the main quantities used in our analysis.

\begin{figure}
    \centering
    \includegraphics[width = 0.495\textwidth]{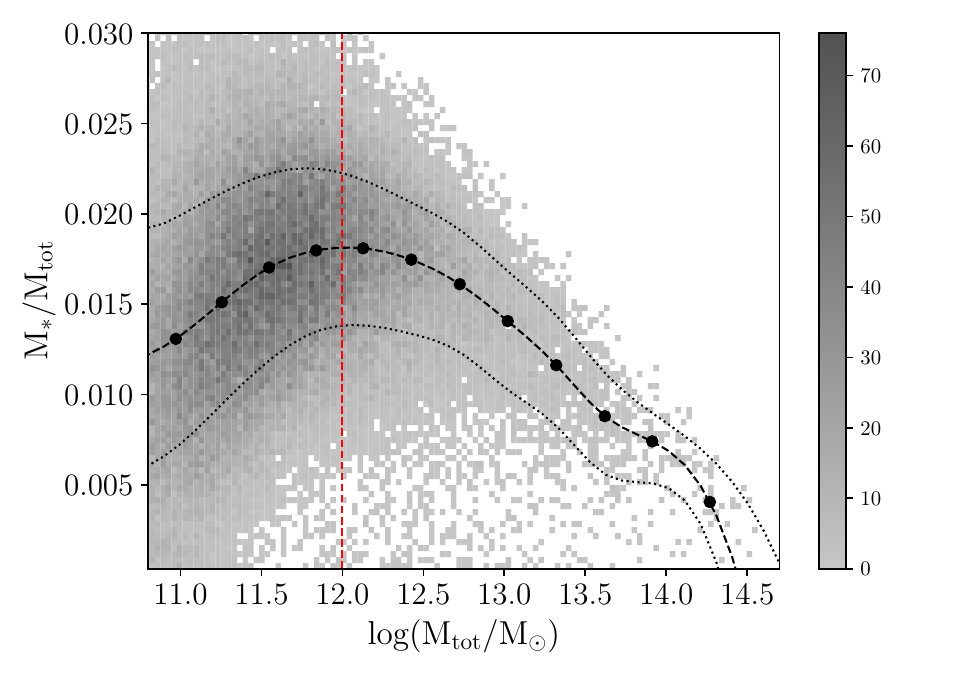}
    
    \caption{2D histogram of $M_{*}/M_{\rm tot}$ versus $M_{\rm tot}$ for central galaxies in the HR5 simulation at a redshift of 0.625. The gray scale bar indicates the number of galaxies in each bin. The black dots and the dotted curved line show the binned median $M_{*}/M_{\rm tot}$ and the interpolated relation, respectively. The dashed curves represent the 16th–84th percentile range of $M_{*}/M_{\rm tot}$ at a fixed $M_{\rm tot}$. Galaxies with $\log(M_{\rm tot}/M_{\odot}) \geq 12$ (marked by the red vertical line) are included in our analysis.}
    \label{fig1}
\end{figure}

\subsection{Sample Selection}

In this study, we choose to work with central galaxies, whose evolution is closely tied with their total mass. These galaxies follow a well-defined relation between their stellar and total mass, unlike satellites, where interactions with more massive neighbors introduce additional complexity. To investigate how the critical mass scale emerges, we focus on galaxies that have already surpassed this critical stage of evolution. A straightforward approach is to select galaxies residing in halos more massive than the total mass at which the stellar-to-total mass ratio peaks. By tracking the evolution of these galaxies back in time, we aim to study the physical changes they undergo as they approach and pass the critical mass.

To determine the stellar mass–total mass relation and define our sample, we selected all central galaxies with $\log(M_{\rm tot}/M_{\odot}) \geq 10.8$ from the final redshift snapshot of the HR5 simulation. This lower total mass cut ensures sufficient resolution for a reliable determination of the median stellar mass–total mass relation in HR5. The distribution of these galaxies at $z_f = 0.625$ is shown as a 2D histogram in the $M_{*}/M_{\rm tot}$ vs. $M_{\rm tot}$ plane in Figure \ref{fig1}. We then binned the galaxies in total mass ($M_{\rm tot}$) bins with widths of 0.3 dex, except for the highest-mass bin, which has a width of 0.6 dex to improve number statistics. The dotted lines in Figure~\ref{fig1} show the 16th and 84th percentiles of the stellar-to-total mass ratio while the dashed line shows the median relation. The relationship between the stellar and total mass components of HR5 galaxies is consistent with previous model predictions and observations \citep{Kim2023}.

Looking at Figure \ref{fig1}, we observe that the ${ M_*/M_{\rm tot}}$ ratio initially increases with halo mass, peaks at ${\log(M_*/M_{\rm tot})}\approx 12$, and then steadily declines with further increases in total mass. Based on this trend, we selected only central galaxies with total masses ${\log(M_*/M_{\rm tot})}\geq 12$ for our analysis. This choice ensures that our sample galaxies have undergone the processes that establish the critical mass scale, making them well suited to investigate its origin. Our final sample includes 19,795 galaxies distributed across eight mass bins.

\begin{figure*}
    \centering

    \includegraphics[width = \textwidth]{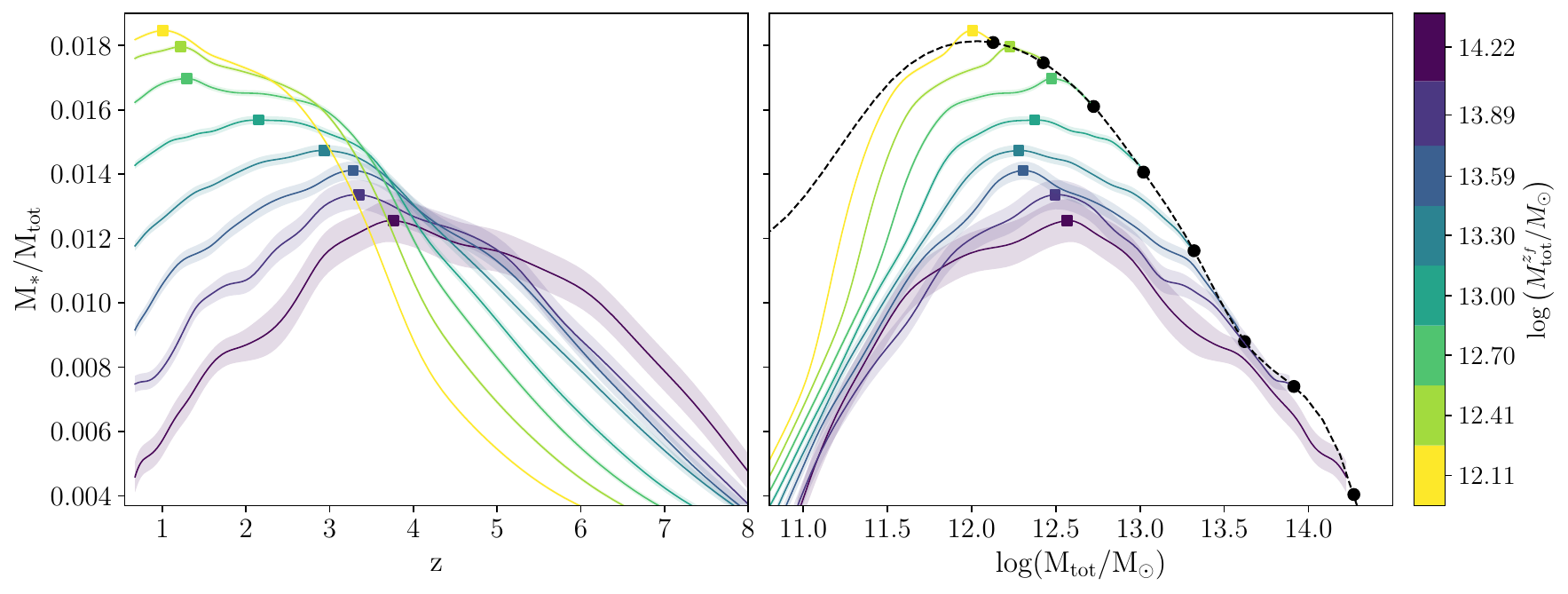}

    \caption{{\bf Left:} The redshift evolution of the median ${ M_*/M_{\rm tot}}$ for galaxies in our sample. {\bf Right:} The evolution of median ${ M_*/M_{\rm tot}}$ as function of total mass of galaxies. Black dots and the dashed line indicate the median stellar mass–total mass relation at the final redshift snapshot of the HR5 simulation, as shown in Figure \ref{fig1}. In both panels, solid lines represent the median evolutionary trajectories of galaxies, color-coded by their final total mass at redshift $z_f =0.625$, as shown by the color bar. The shaded region enclosing the lines is standard error on the median. The galaxy total mass and redshift corresponding to the peak $M_{*}/M_{\rm tot}$ are marked with boxes.} 
    \label{fig2}

\end{figure*}

\subsection{Evolutionary History}

To study the evolutionary history of galaxies, it is essential to identify their progenitors. The progenitor–descendant relationship across HR5 snapshots is constructed using merger trees, as described in \cite{Park_2022} and \cite{Lee2021}. The main progenitor is defined as the most massive progenitor in the merger tree from the previous snapshot. In this work, we track galaxies along the main progenitor branch.

The evolutionary trajectories of individual galaxies can differ greatly. Since we are interested in grasping the general evolutionary trend, we first group our sample galaxies according to their total mass.
Then to construct a median evolutionary trend of a certain property, such as the stellar-to-halo mass ratio (STR), $M_{*}/M_{\rm tot}$, we take the following steps:

\begin{enumerate} 
\item Select all galaxies belonging to a specific total mass bin at the lowest redshift $z_f$
\item Trace the redshift evolution of the chosen physical property (ex. $M_{*}/M_{\rm tot}$) for each selected galaxy across all snapshots up to redshift $z = 8.5$ and compute its median value along with the associated standard error estimated as $1.25\,\sigma/\sqrt{N}$ \citep[see eg.][]{KendallStuart1977}.

\end{enumerate}

These steps yield the representative median evolutionary history of the chosen physical properties of galaxies.

Since the redshift evolution of galaxy properties and their median trends fluctuate significantly, we apply Gaussian smoothing over a constant number of snapshots, corresponding to an effective temporal smoothing of $100\mbox{--}400$\,Myr. 
 We note that the progenitors of the lowest-mass galaxies in our sample reach a total mass of $\log(M_{\rm tot}/M_{\odot}) \approx 10.4$ at the highest redshift $z = 8.5$, which sets the lowest mass limit of our analysis. From this point onward, we label galaxies and their progenitors based on the total mass of the galaxies at the final redshift snapshot.

\section{Results} \label{sec:results}

\subsection{Stellar-to-Total Mass Ratio}

We now present our results on the redshift evolution of the median STR for galaxies with different $M_{\rm tot}(z_f)$. This evolutionary trend is shown in the left panel of Figure \ref{fig2}. Although the STR evolution differs for different final masses, a general trend can be observed: the STR first monotonically increases, reaches a peak, and then declines steadily with time. At very high redshifts ($z > 6$), progenitors of more massive galaxies tend to have higher STR values. They also reach a peak relatively earlier in time. Progenitors of lower-mass galaxies, even though they start with a lower value of STR, quickly catch up and surpass the high-mass galaxies in STR. This results in a trend where the peak STR is inversely related to the galaxy mass at the final snapshot, i.e., the peak STR achieved by progenitors of lower-mass galaxies is higher than that achieved by progenitors of massive galaxies. The relation is well-approximated by a linear relation, $(M_*/M_{\rm tot})_{\rm peak} = -0.003 \log (M_{\rm tot}({z_f})/M_{\odot}) +0.053$. 

In the right panel of Figure \ref{fig2}, we plot the trajectories of progenitor galaxies in the $M_{*}/M_{\rm tot}$ vs. $\log(M_{\rm tot}/M_{\odot}$) plane. A striking feature of this plot is that, regardless of the final total mass (circular dots), the total mass of galaxies at which the STR peaks (squares) lies within a narrow range of $12.1 \leq {\log(M_{\rm tot}/M_{\odot})} \leq 12.6$.

\begin{figure*}
  \centering
  \includegraphics[width = 0.85\textwidth]{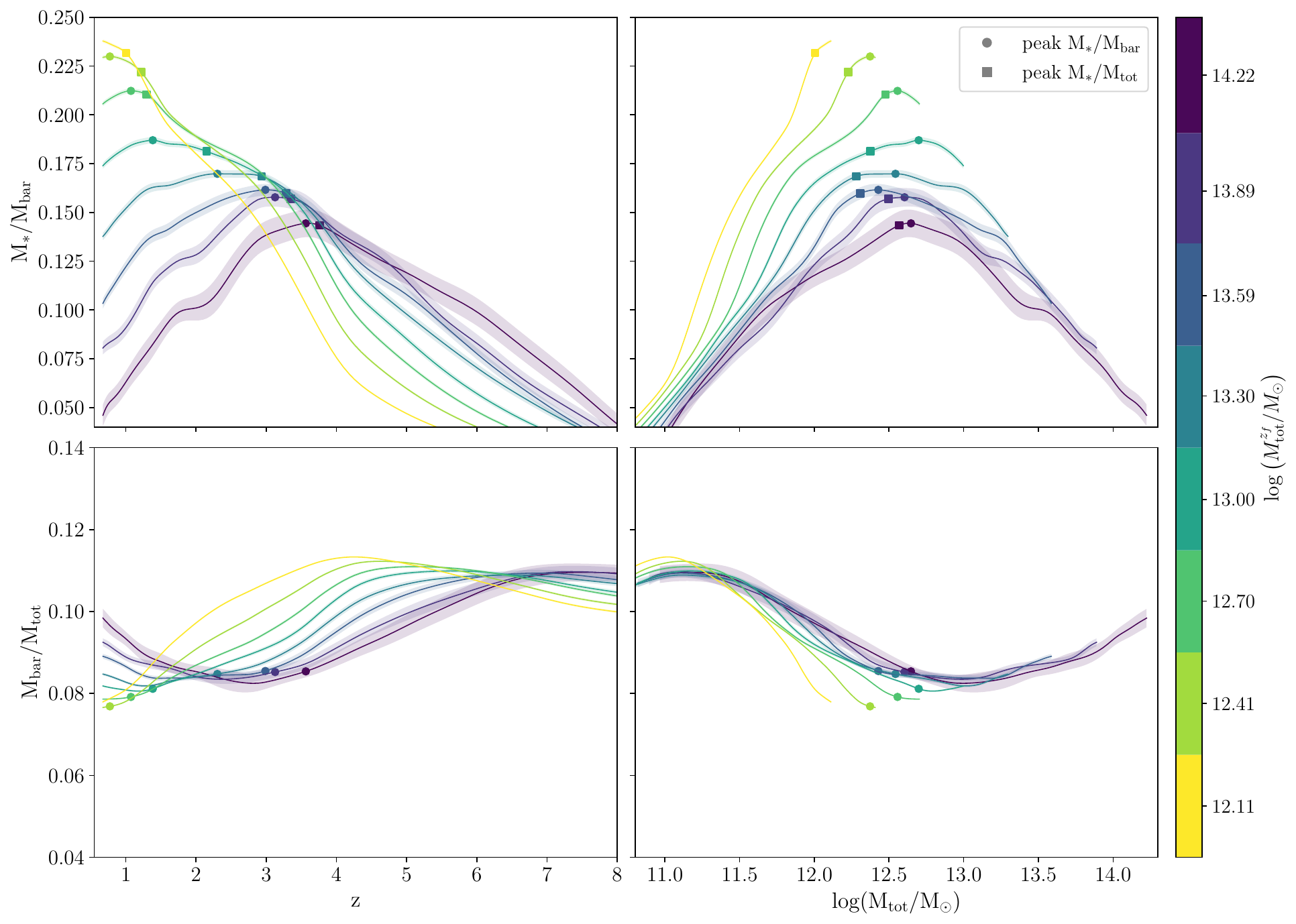}
  \caption{The evolution of SBR as a function of redshift (top left) and total mass (top right), and the evolution of BRF as a function of redshift (bottom left) and total mass (bottom right). The colored solid lines and shaded region has same meaning as in previous figure. The circle and square markers indicate the redshift and total mass at which galaxies reached their peak ${\rm M_/M_{bar}}$ and peak ${\rm M_/M_{tot}}$, respectively, during their evolution.}
  \label{fig3}
\end{figure*}

\subsection{Stellar-to-baryon Mass Ratio \& Baryon Retention Fraction}

The existence of a narrow range of galaxy mass at which the STR begins to decline is intriguing and may indicate that, upon reaching this mass scale, the physical processes governing mass growth begin to change. The STR is determined not only by star formation efficiency, but also by the amount of baryons acquired via accretion and mergers, modulated by feedback processes, and ultimately retained by galaxies.

To disentangle these effects, we express the STR as the product of two quantities:
\begin{equation}
\label{eq:fig1}
{M_*/M_{\rm tot}=M_*/M_{\rm bar} \times M_{\rm bar}/M_{\rm tot}}
\end{equation}
Here, ${M_{\rm bar}}$ represents the total baryonic mass in the galaxy, which is the sum of the gas mass (${M_{\rm gas}}$) and the stellar mass (${ M_*}$). The first term, 
${M_*/M_{\rm bar}}$, is the stellar-to-baryon mass ratio (SBR), and we call the second term, ${M_{\rm bar}/M_{\rm tot}}$ as baryon retention fraction (BRF). We now separately examine the evolution of SBR and BRF of our sample galaxies. 
 
The top-left panel of Figure \ref{fig3} shows how SBR evolves with redshift. It initially increases with cosmic time, reaches a peak, and then declines steadily. Note that for the least massive galaxies in our sample, the SBR peak occur at final redshift and may not correspond with eventual turnover. Strikingly, when examining the evolution of SBR as a function of total galaxy mass (top-right panel of Figure \ref{fig3}), we find that galaxies reach their peak SBR (circular dots) within a remarkably narrow mass range: $12.37 \leq {\rm log(M_{\rm tot}/M_{\odot})} \leq 12.69$. 

The qualitative nature of SBR evolution closely resembles that of STR. In the top panels of Figure \ref{fig3}, we compare the epochs and total masses at which galaxies reach their peak STR and SBR. We find that the STR peak precedes the SBR peak in cosmic time with the difference being at most 0.3 dex in total mass.

The bottom-left panel of Figure \ref{fig3} displays the redshift evolution of the median BRF. The BRF shows a mild increase, followed by a gradual decline up to the redshift corresponding to the peak SBR. Beyond this point, the BRF gradually begins to increase monotonically. 

We see from Figure \ref{fig3} that, during the evolutionary history of galaxies, the BRF changes by only about $\sim 30\%$, whereas the SBR changes by a factor of $\sim 3$–$4$. This indicates that the SBR is the dominant factor determining the evolution of the STR. It is then natural to ask: what causes the SBR of galaxies to decline once they reach a certain critical mass?

To better understand the turnover of ${M_{*}/M_{\rm bar}}$, we need to examine the gas content of galaxies and its conversion into stars. As the fuel for star formation, gas plays a central role in regulating stellar mass growth in galaxies. The balance between gas consumption through efficient star formation and gas removal or heating by star formation–driven feedback governs the available gas reservoir, and thus shapes the efficiency with which galaxies convert baryons into stars. We address these processes in detail in the following subsections.

\subsection{Evolution of total gas and ISM mass fractions}

\begin{figure}
  \centering
  \includegraphics[width=0.96\columnwidth]{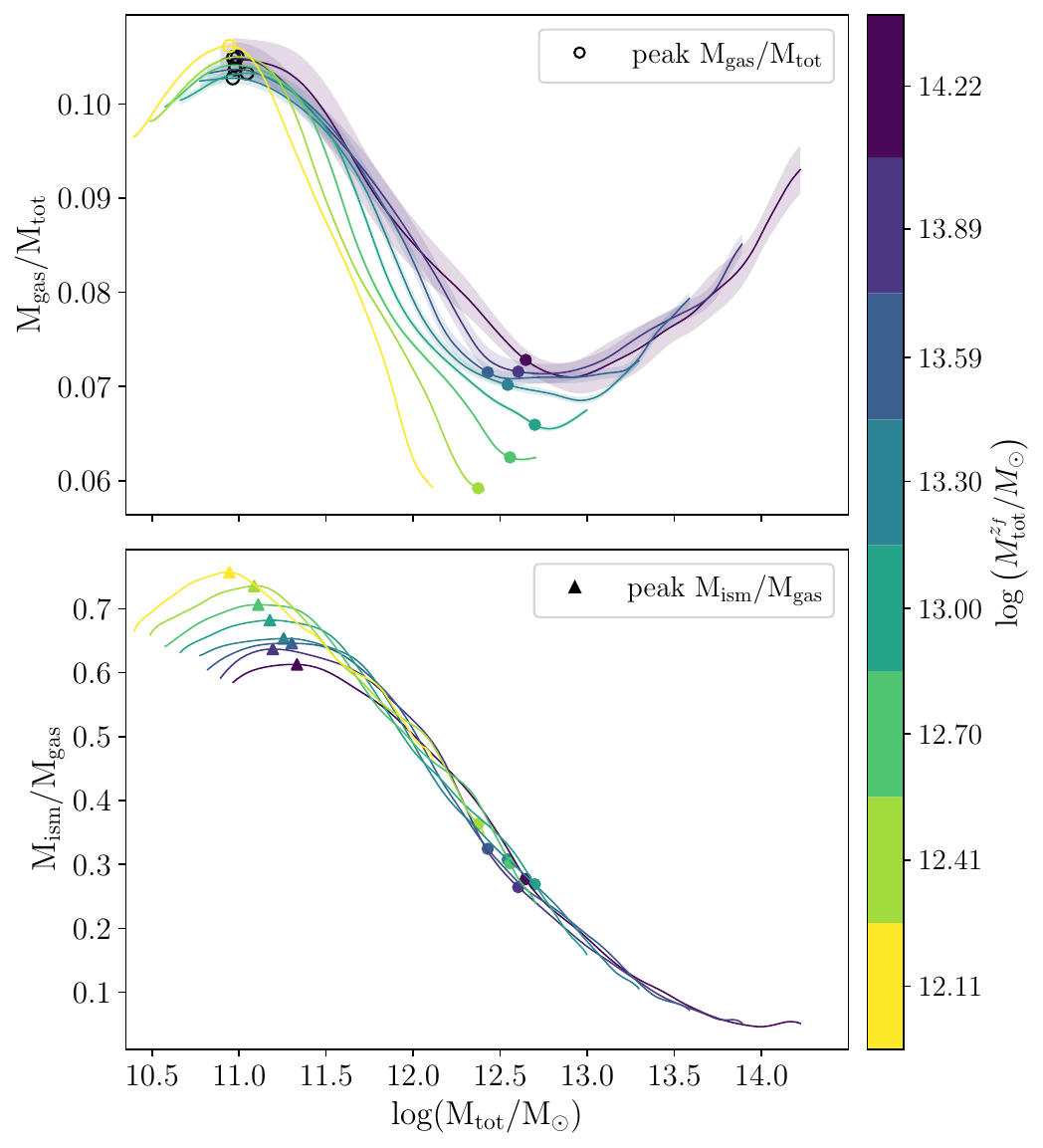}
  \caption{{\bf Top:} The evolution of gas retention fraction of galaxies as function of total mass. {\bf Bottom:} The evolution of fraction of gas mass available as the ISM of galaxies. The colored lines and shaded regions have the same meaning as in the previous figures. The halo mass corresponding to the peak gas retention fraction ratio and peak ISM fraction are marked by open circles and triangles, respectively.}
  \label{fig4}
\end{figure}

To characterize the gas content of galaxies and its evolution, in Figure \ref{fig4} we present the evolution of ${ M_{\rm gas}/M_{\rm tot}}$ and ${ M_{\rm ISM}/M_{\rm gas}}$ as a function of the total galaxy mass $M_{\rm tot}$. Here, $M_{\rm ISM}$ is defined as the gas enclosed within five times the stellar half-mass radius of galaxies while $M_{\rm gas}$ is the total gas mass in galaxies.

Looking at the top panel of Figure \ref{fig4}, we find that the gas retention fraction $M_{\rm gas}/M_{\rm tot}$ peaks at the first critical mass scale of $M_{\rm tot} \approx 10^{11} M_{\odot}$ for all galaxies, after which it begins to decline. The fraction continues to decrease with increasing  $M_{\rm tot}$ until galaxies reach the second critical mass scale associated with the SBR turnover. Beyond this critical mass, $M_{\rm gas}/M_{\rm tot}$ begins to increase.

We use the ISM fraction $M_{\rm ISM}/M_{\rm gas}$ to represent the fraction of gas mass contained within the main optical extent of galaxies. The bottom panel of Figure \ref{fig4} shows that galaxies begin their early evolutionary phase with most of their gas residing in the ISM. The peak ISM fraction is lower and occurs at higher mass for progenitors of more massive galaxies. After the peak, the ISM fraction decreases steeply, and at the critical mass scale associated with the peak SBR, only about 30\% of the gas remains in the ISM phase.

In understanding the processes driving the evolution of the gas retention fraction, one might expect feedback processes to be the primary factor modulating gas retention in galaxies. However, at any given time, $M_{\rm gas}$ denotes the amount of gas present within the full extent of galaxies, without accounting for the gas that has been expelled or converted into stars. In other words, the retained $M_{\rm gas}$ satisfies $M_{\rm gas} = M^{\rm initial}_{\rm gas} + M^{\rm accreted}_{\rm gas} - M^{\rm ejected}_{\rm gas} + M^{\rm recycled}_{\rm gas} - M^{\rm SF}_{\rm gas}$. Although ejective feedback may appear as the main mechanism regulating the gas content, the last term—gas consumed by star formation—is also important, not only because it directly depletes the available gas reservoir, but also because increased star formation strengthens feedback, leading to further gas expulsion. In fact, weakened in-situ star formation will simultaneously slow the growth rate of stellar mass while increasing gas retention in galaxies, thereby causing the SBR to decline. Thus, in-situ star formation directly regulates gas retention. While AGN feedback may contribute by affecting star formation in high-mass galaxies \citep[see eg.][]{Dubois2016}, its role is indirect since black hole growth is itself regulated by supernova feedback and cumulative mass assembly \citep{Booth2010, Booth2011, Dubois2015, Habouzit2017}. We also find that peak AGN activity does not coincide with the critical mass where the gas retention fraction turns over (Appendix~\ref{AGNrate}). Given the role of in-situ star formation as the immediate regulator of gas retention, we next examine its impact on the evolution of the SBR.

\subsection{In-situ Star Formation and the Critical Mass}

\begin{figure}
    \centering
    \includegraphics[width = 0.495\textwidth]{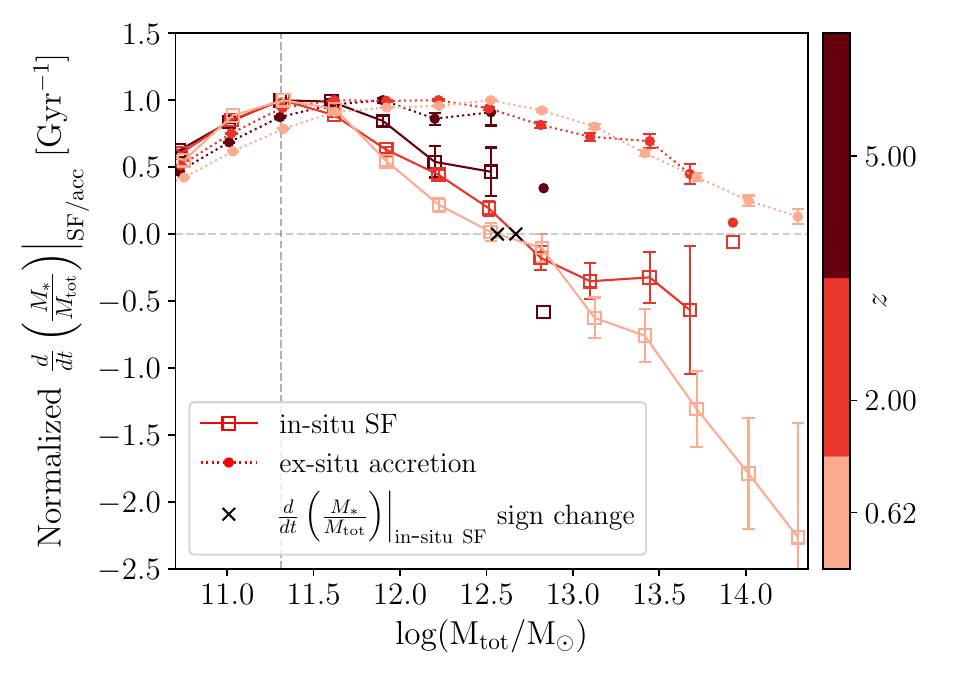}
    \caption{Normalized median rate of change of the stellar-to-total mass ratio, $\frac{d}{dt}\left(\frac{M_*}{M_{\rm tot}}\right)$, as a function of the total mass of central galaxies. Contributions from in-situ star formation and ex-situ accretion are shown by open boxes (solid lines) and filled circles (dotted lines), respectively; error bars indicate the uncertainty on the median. Owing to the small number of galaxies in the highest-mass bins, error bars are omitted and the points are not joined to adjacent bins. The colors represent redshifts $z = 5$, 2, and 0.625. The vertical gray lines mark the mass scale where the in-situ contribution peaks. The two black crosses mark the points where $\frac{d}{dt}\left(\frac{M_*}{M_{\rm tot}}\right)|_{\rm in\text{-}situ\,SF}$ becomes zero and changes sign thereafter}
    \label{fig5}
\end{figure}

In this subsection, we focus on how stellar mass growth due to in-situ star formation affects the gas retention fraction ${M_{\rm gas}/M_{\rm tot}} = {M_{\rm bar}/M_{\rm tot}} - {M_{*}/M_{\rm tot}}$. It should be noted that stellar mass growth (and hence ${M_{*}/M_{\rm tot}}$ evolution) occurs through both in-situ star formation and ex-situ accretion. Only in-situ growth depletes the gas reservoir by converting gas into stars, while ex-situ growth adds pre-existing stars via accretion and therefore does not reduce the gas content. Furthermore, baryon retention $M_{\rm bar}/M_{\rm tot}$ itself is not constant and is affected by accretion and ejective feedback associated with in-situ star formation.

In Figure \ref{fig5}, we plot the peak-normalised median rate of change of STR, i.e.~$\frac{d}{dt}\left(\frac{M_{*}}{M_{\rm tot}}\right)$, due to stellar mass growth through in-situ star formation (solid lines) and ex-situ accretion channels (dashed lines), as a function of $M_{\rm tot}(z)$ of galaxies at $z=5, 2$, and 0.625. To estimate it, we first separately compute the increase in stellar mass ($\Delta M_{*}^{\rm SF/acc}$) of a galaxy due to in-situ star formation or accretion between two consecutive snapshots. The quantity $\frac{d}{dt}\left(\frac{M_*}{M_{\rm tot}}\right)|_{\rm SF/acc}$ at snapshot $n$ (or a redshift) is then computed as $\frac{1}{\Delta t}\left(\frac{M_{*,n-1}+\Delta M_*^{\rm SF / accn}}{M_{{\rm tot},n}}-\frac{M_{*,n-1}}{M_{{\rm tot},n-1}}\right)$, where $\Delta t$ is the time separation between snapshots $n-1$ and $n$. The peak absolute rate of change of STR due to in-situ star formation at these redshifts are 0.0211, 0.0045, and 0.0007, decreasing by orders of magnitude over time; we therefore normalize the curves to compare their relative trends. For comparison, we also plot the normalized median ex-situ contribution to this rate. Its absolute peak increases with time, reaching 0.046, 0.029, and 0.014 at $z =$ 5, 2 and 0.625 respectively.

Examining Figure \ref{fig5}, we find that the median rate of change of the STR due
to in-situ star formation increases with total mass and peaks at
$\sim 10^{11.3}\,{\rm M_\odot}$. Beyond this mass, it declines with increasing
total mass and eventually becomes negative. The total mass at which
$\frac{d}{dt}\left(\frac{M_*}{M_{\rm tot}}\right)|_{\rm SF}$
transitions from positive to negative is $\sim 10^{12.67}\,{\rm M_\odot}$ and
$\sim 10^{12.56}\,{\rm M_\odot}$ at redshifts $z = 2$ and 0.625, respectively.
For the $z = 5$ snapshot, the transition mass is similar but uncertain due to small number of galaxies.

A positive $\frac{d}{dt}\left(\frac{M_*}{M_{\rm tot}}\right)|_{\rm SF}$ indicates efficient conversion of gas into stars, steadily lowering $M_{\rm gas}/M_{\rm tot}$. 
Once galaxies cross the critical mass scale of $\sim 10^{12.5}\,{\rm M_\odot}$, however, the rate of change of STR becomes negative, signalling that in-situ star formation has become inefficient to the point that total mass growth exceeds in-situ stellar mass growth. Gas is then no longer efficiently consumed or expelled and begins to accumulate, producing an upturn in $M_{\rm gas}/M_{\rm tot}$. The enhanced accumulation of gas, coupled with the slower growth of stellar mass, drives the decline of the SBR at the critical mass scale.

\subsection{Gas Accretion, Hot Gas Fraction, and Cooling}

\begin{figure}
  \centering
  \includegraphics[width=0.96\columnwidth]{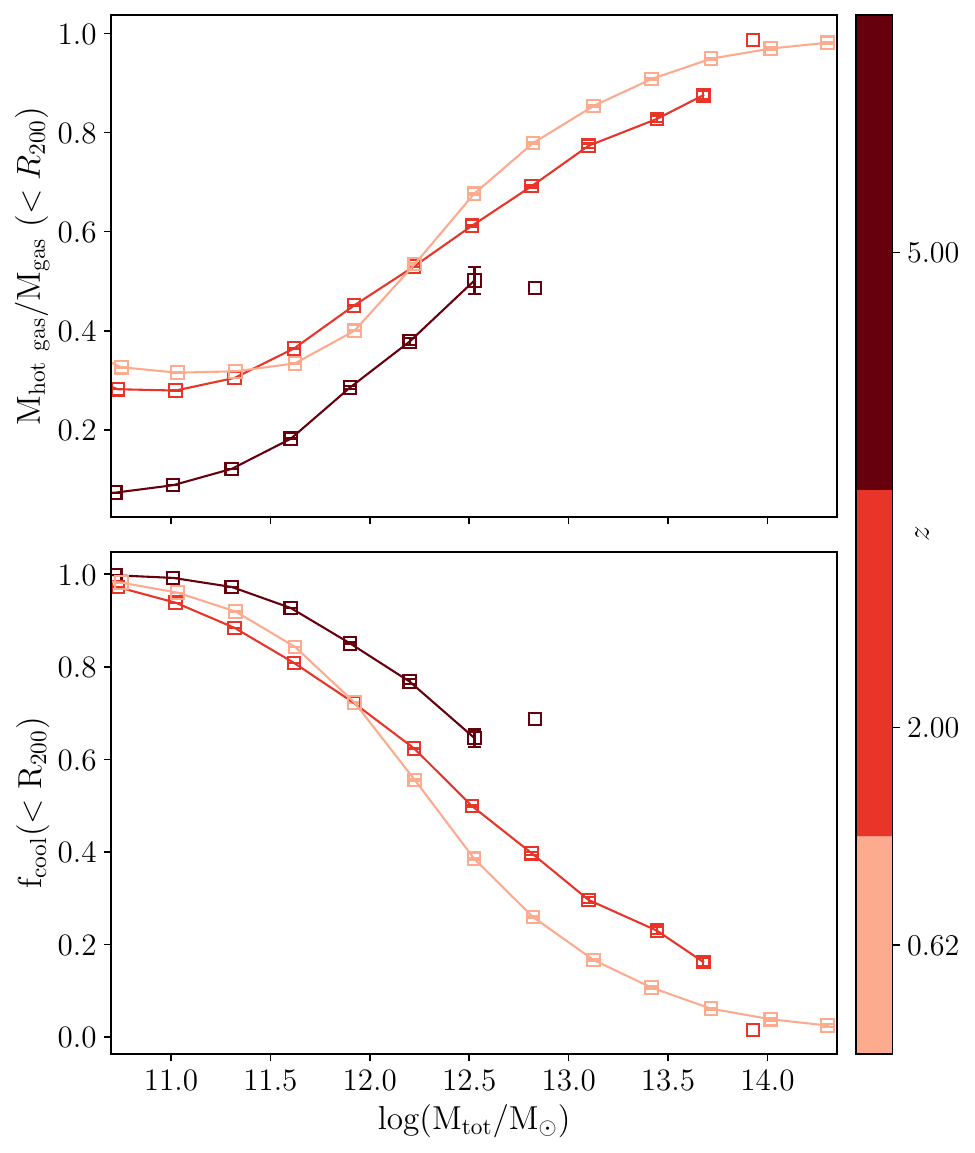}
  \caption{{\bf Top:} The median hot gas mass fraction of galaxies as a function of total mass. {\bf Bottom:} The median fraction of gas able to cool within the dynamical timescale as a function of total mass. In both panels, error bars indicate the uncertainty on the median, and colors represent the redshifts $z = 5$, 2, and 0.625. Error bars and connecting lines are omitted for the highest-mass bins due to low number statistics.}
  \label{fig6}
\end{figure}

The previous discussion highlights the coincidence between the inefficient in-situ star formation and the establishment of the critical mass scale of $\sim 10^{12.5} \ {\rm M_{\odot}}$. 
We now investigate what physical processes lead to this inefficiency around this mass scale. We note that for sustained star formation a steady supply of cool, dense gas is required. However, previous studies have shown that galaxies tend to develop a hot gas atmosphere as they grow in mass over time \citep[see eg.][]{Keres2005, Dekel_2006, Ocvirk2008, Correa2018}. This heating is driven by virial accretion shocks as well as feedback from supernovae and AGN. Once the halo becomes sufficiently massive, the heated gas remains gravitationally bound to the system, forming a stable hot gas halo that suppresses further cooling and star formation.

To test this scenario, we measure two quantities. The first is the hot-gas fraction, defined as the ratio of hot-gas mass to total gas mass within a sphere of radius ${\rm R_{200}}$, where $R_{200}$ is the radius within which the average matter density is 200 times the critical density of the universe, and “hot gas” refers to gas with temperature greater than $3\times10^4$ K. The second quantity, denoted as $f_{\rm cool}$, is the mass fraction of gas whose cooling timescale is shorter than the dynamical timescale within ${\rm R_{200}}$, 
\begin{equation}
{f_{\rm cool}=M_{\rm gas}(\tau_{\rm cool} < \tau_{\rm dyn}; <R_{200}) / M_{\rm gas}(<R_{200}).}
\end{equation}
The cooling timescale $\tau_{\rm cool}$ is computed using gas temperature and metallicity based on the cooling function of \cite{Sutherland1993}. The dynamical timescale $\tau_{\rm dyn}$ is estimated as the ratio of $R_{200}$ to $V_{200}$, where $V_{200} = \left[ G M(<R_{200})/R_{200} \right]^{1/2}$. We plot the median trends of the hot-gas fraction and $f_{\rm cool}$ for central galaxies at redshifts $z = 5$, $2$, and $0.625$ in Figure~\ref{fig6}.

\begin{figure*}
    \centering

    \includegraphics[width = 0.495\textwidth]{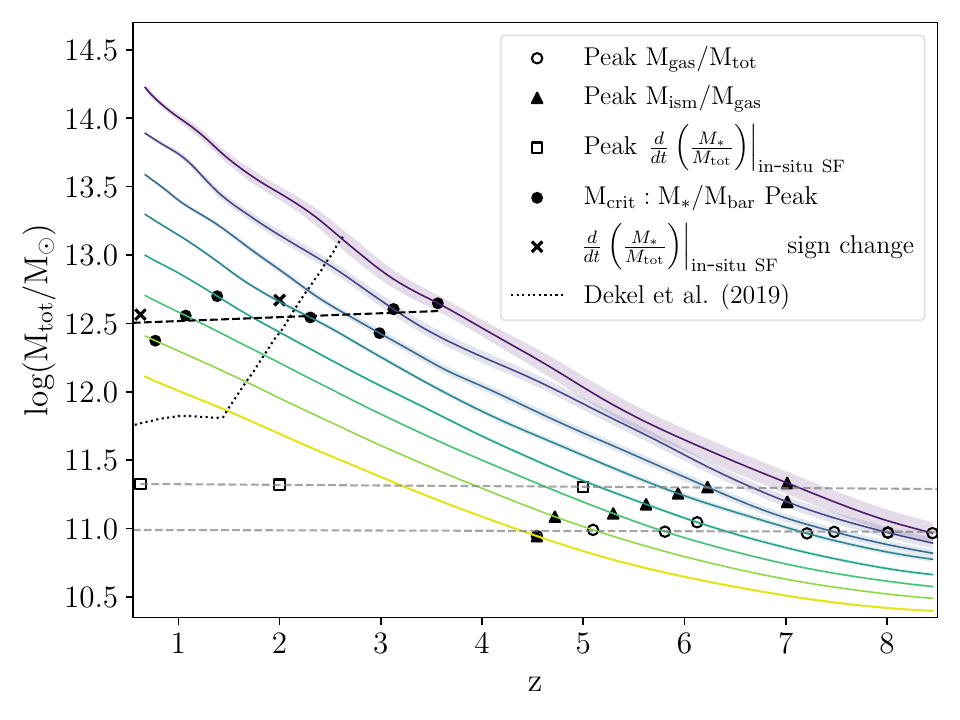}
    \includegraphics[width = 0.495\textwidth]{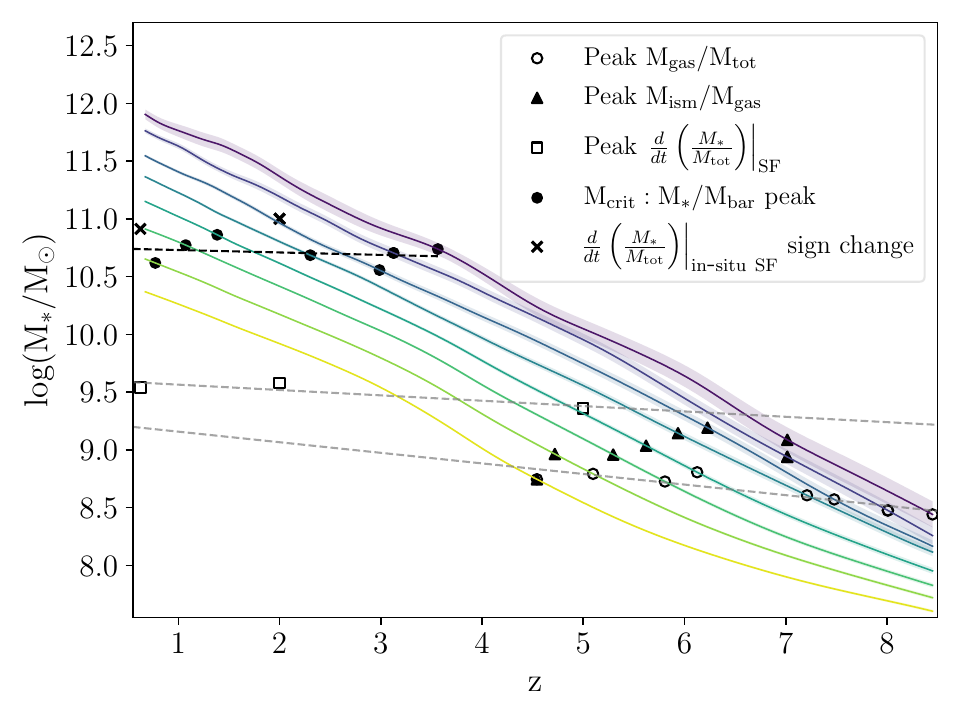}

    \caption{{\bf Left:} Redshift evolution of the median total mass of galaxies, marked with mass scale associated important transitions. The solid line and shade regions denote median mass and its uncertainty. The open circle, filled triangles, and open squares indicate mass scale associated with peak $M_{\rm gas}/M_{\rm tot}$, $M_{\rm ISM}/M_{\rm gas}$ and $\frac{d}{dt}\left(\frac{M_*}{M_{\rm tot}}\right)|_{\rm in\text{-}situ~SF}$ respectively. Grey dashed lines show linear fits to the open circles and squares. Filled circles denote the critical mass scale where the stellar-to-baryon mass ratio (SBR) peaks, and dashed black line is a linear fit. The black dotted line shows the critical-mass prediction from \citet{Dekel_2019}. {\bf Right:} Redshift evolution of the median stellar mass of galaxies. Symbols follow the same conventions as in the left panel.}

    \label{fig7}
\end{figure*}

The upper panel of Figure \ref{fig6} shows that the hot gas mass fraction monotonically increases with $M_{\rm tot}$. This fraction does not change much with increasing mass up to a total mass of $\sim 10^{11.3-11.7}\ {\rm M_{\odot}}$. Above this mass threshold, the hot gas fraction increases rapidly, and by the time they reach the critical mass scale of $\sim 10^{12.5}\ {\rm M_{\odot}}$, $M_{\rm hot\ gas}/M_{\rm gas}(<R_{200})$  rises most steeply and the majority of the gas is in the hot phase.

Further useful insights into the nature of gas accretion are gained by examining the trend of ${f_{\rm cool}}$. The quantity ${f_{\rm cool}}$ indicates the mass fraction of gas that is able to cool before falling onto the ISM of galaxies. We find in the bottom panel of Figure \ref{fig6} that low-mass galaxies have a high value of ${f_{\rm cool}}$, with a slow decline up to a total mass of $\sim 10^{11.3}\ {\rm M_{\odot}}$, beyond which ${ f_{\rm cool}}$ declines rapidly and the majority of the gas cannot cool within the dynamical time scale above the critical mass. 

This result suggests the onset of a dynamically stable hot halo that reduces the gas supply to the ISM, explaining the declining ISM fraction seen in the bottom panel of Figure \ref{fig4}. The establishment of this hot halo suppresses in-situ star formation, which in turn increases the gas retention fraction through reduced gas consumption and weaker stellar feedback. Together, these effects reduce in-situ stellar mass growth while baryonic mass continues to grow, producing downturns in both SBR and STR beyond the critical mass.

\subsection{Critical Mass and its redshift dependence}

In this subsection, we synthesize the above findings to construct a simple evolutionary picture of galaxy growth, highlighting the key phases and the emergence of the critical mass.

Previous sections show that a stable hot gas halo is well-established at $\log (M_{\rm tot}/M_{\odot}) \approx 12.5$ and the star formation becomes very inefficient as indicated by a sign change in $\frac{d}{dt}\left(\frac{M_*}{M_{\rm tot}}\right)|_{\rm SF}$ from positive to negative.  Consequently, galaxies pass through a critical evolutionary phase, characterized by an upturn in $M_{\rm gas}/M_{\rm tot}$, and downturns in both $M_*/M_{\rm bar}$ and $M_*/M_{\rm tot}$.
In particular, the peak in the SBR occurs within a very narrow mass range for all galaxies with different final total mass.

We define the critical mass threshold as the total mass of a galaxy where a stable hot gas halo is established as indicated by a peak and downturn of the SBR. This definition provides a natural marker for the transition in gas and baryon growth and the onset of inefficient in-situ star formation. Furthermore, since the SBR largely governs the evolution of the STR, galaxies that surpass this critical mass scale are expected to follow a declining trend in their STR.

In the left panel of Figure \ref{fig7}, we show the evolutionary history of total mass growth in galaxies and mark the key events that shape their evolution. The overall picture is as follows. Galaxies grow their total mass monotonically with time. When they reach a total mass of $\sim10^{11}\ {\rm M_{\odot}}$, their gas retention fraction peaks to $M_{\rm gas}/M_{\rm tot} \approx 0.104$ (open circles). Above this first critical mass $M_{\rm gas}/M_{\rm tot}$ begins to decline due to a combination of more efficient gas consumption from star formation and removal of gas through feedback processes. Shortly after this, continued gas heating caused by feedback and accretion shocks makes the gas increasingly stable against collapse toward the galaxy center. This reduces the supply of gas to the ISM, leading to a decline in the ISM fraction (black filled triangles). The reduced ISM supply begins to affect in-situ star formation. The in-situ star formation efficiency peaks around $\sim10^{11.3}\ {\rm M_{\odot}}$ (open squares) and then gradually declines. Despite this decline, star formation remains active enough to continue consuming gas, further lowering the total gas mass fraction. During this stage, stellar mass continues to grow while baryonic mass growth slows, causing the SBR to increase. 

By the time galaxies reach their critical mass, the hot dynamically supported gas halo appears to be fully developed. This halo cuts off the supply of cool gas to the ISM, making in-situ star formation highly inefficient. With star formation no longer able to efficiently consume gas, the hot gas reservoir begins to build up, leading to an increase in the total gas mass fraction. The combination of slower stellar mass growth and enhanced baryonic mass growth ultimately causes the SBR to turn over at the critical mass (black filled circles). The median rate of change of the STR also changes sign from positive to negative at similar mass scale (black crosses). We do not find any definite trend of the critical mass with redshift; a linear fit results in a relation $\log\left(M_{\rm tot}^{\rm crit}/{\rm M_\odot}\right)= 0.029\,z + 12.488$. The critical mass appears to be roughly independent of redshift, with a median value of $\sim 10^{12.5}{\rm M_{\odot}}$.

For comparison, we also show the critical mass predicted by \citet{Dekel_2019}, associated with galaxy quenching, black hole growth, and the turnover of the stellar mass–total mass relation in semi-analytical models based on \citet{Dekel_2006}. They predict a nearly constant critical mass below $z \sim 1.5$ at $\approx 10^{11.8}{\rm M_{\odot}}$, which increases steeply at higher redshift. This trend is not seen in our results and may arise from differences in the treatment of baryonic physics and feedback, particularly gas inflow and retention, between their semi-analytical models and the HR5 simulation.

In the right panel of Figure \ref{fig7}, we present the median stellar mass evolution of galaxies as a function of redshift. As before, all important events in the evolutionary history of galaxies have been marked including the critical stellar mass. The redshift dependence of critical stellar mass can be described by a linear relation $\log\left(M_*/{\rm M_\odot}\right) = -0.021\,z + 10.751$. We find the critical stellar mass to be also roughly independent of redshift, with a median value of ${\rm 10^{10.7} M_{\odot}}$, consistent with the stellar mass scale associated with various transitions in galaxy properties outlined in the introduction.

We also find that galaxies undergo several evolutionary changes within a total
mass range of $\sim 10^{11\text{-}11.3}\,{\rm M_\odot}$, corresponding to a
stellar-mass range of $\sim 10^{8.5\text{-}9.5}\,{\rm M_\odot}$. 
First, the gas retention fraction $M_{\rm gas}/M_{\rm tot}$ peaks at $ 10^{11}{\rm M_\odot}$. We refer to this mass as the secondary critical mass scale. Above the mass and within the mass range the ISM fraction reaches the maximum value, the median rate of increase of the STR becomes highest due to efficient in-situ star formation, and the majority of the gas resides in the cool component. The galaxies less massive than this secondary critical mass scale seems to evolve differently. Interestingly, this mass scale is also consistent with several transitions reported in the literature, including a change in the slope of the stellar mass-star formation rate relation at $\sim 10^{9.5}\,{\rm M_\odot}$ \citep[e.g.,][]{Huang2012}, which remains approximately redshift-independent out to $z \sim 5$ \citep{Merida2025}. Breaks near this mass are also seen in the stellar mass-metallicity relation \citep{Blanc2019, Curtis2024, Raptis2025}, the galaxy half-light radius-stellar mass relation \citep[see for eg.][]{Roy2018, Mishra2023}, and in the transition between thick- and thin-disk--dominated galaxies \citep[see eg.][]{Tsukui2024}.

We note that our result of a redshift-independent critical mass ($M^{\rm crit}_{\rm tot}\sim10^{12.5}{\rm M_\odot}$ or $M^{\rm crit}_{\rm *}\sim10^{10.7}{\rm M_\odot}$), associated with the turnover of the SBR and STR, the minimum and subsequent upturn of the gas fraction, and the formation of hot gas halo, is broadly consistent with other state-of-the-art simulations that adopt different astrophysical models and feedback prescriptions from HR5. Using IllustrisTNG suite of CAMELS simulation with fiducial feedback parameters, \citet{Tortora2025} find a critical stellar mass corresponding to the peak stellar-to-dark matter mass ratio of $\sim 10^{10.6}{\rm M_\odot}$ at $z=0$, evolving weakly to $\sim 10^{10.75}{\rm M_\odot}$ by $z\sim2$. Similarly, \citet{LucieSmith2025} find in the FLAMINGO simulation a redshift-independent critical mass of $M_{200}\sim 10^{12.8}{\rm M_\odot}$ at which $M_{\rm gas}/M_{\rm tot}$ reaches a minimum, consistent with our gas fraction results (see eg. Figure~\ref{fig4}). They also show that while the absolute values of $M_{\rm gas}/M_{\rm tot}$ change with different feedback prescriptions, the value of critical $M_{200}$ at which the minimum occurs roughly remains same. Finally, \citet{Correa2018}, using the EAGLE simulation, find that galaxies with total mass $\sim 10^{11.5-12}{\rm M_\odot}$ develop a hot gas halo independent of redshift. They define the hot halo as being sufficiently established when $\gtrsim 10\%$ of the gas satisfies $\tau_{\rm cool} > \tau_{\rm dyn}$, which is analogous to $1 - f_{\rm cool}$ in our definition. From Figure \ref{fig6}, we find that the equivalent criterion ($1 - f_{\rm cool} \gtrsim 0.1$) is met at total galaxy masses of $\sim 10^{11.5-12}{\rm M_\odot}$, consistent with \citet{Correa2018}. Our results on the hot gas mass fraction are also consistent with X-ray observational constraints (see Appendix~\ref{obs}). Collectively, these results suggest that the critical mass scale is a common feature of galaxy evolution that consistently emerges across different simulations, despite variations in feedback implementations and astrophysical modeling.

\section{Summary}\label{sec:summary}

We investigate the physical origin of the critical mass scale, a threshold where galaxy properties and scaling relations undergo fundamental transitions, using the Horizon Run 5 cosmological hydrodynamical simulation. We focus on the mass scale associated with the turnover of the stellar-to-total mass ratio (STR) in central galaxies as a function of total mass. By tracing the evolutionary histories of massive ($M_{\rm tot} \geq 10^{12} {\rm M_\odot}$) central galaxies, we identify the key physical processes driving this turnover and construct a coherent picture for the emergence of the critical mass scale. Our main results are summarized below:

$\bullet$ We find that a redshift-independent critical mass scale at $\sim 10^{12.5}\ {\rm M_{\odot}}$ in total mass (or $\sim 10^{10.7}\ {\rm M_{\odot}}$ in stellar mass) naturally arises from the interplay between the baryon cycle and star-formation efficiency. This threshold marks a phase with an upturn in gas fraction ($M_{\rm gas}/M_{\rm tot}$) and baryon retention fraction (BRF, $M_{\rm bar}/M_{\rm tot}$), and a downturn in the stellar-to-baryon mass ratio (SBR, $M_{\rm *}/M_{\rm bar}$), driven by the formation of a dynamically stable hot gas halo.

$\bullet$ At this critical mass, the hot gas halo strongly suppresses cool gas inflow, reducing in-situ star formation efficiency to the point that total mass growth exceeds in-situ stellar mass growth. As gas is no longer efficiently consumed, the hot gas reservoir grows while stellar mass growth slows, producing the observed increase in gas fraction and BRF and the turnover in SBR.

$\bullet$ We decompose STR as the product of SBR and BRF and find that its evolution is dominated by SBR. BRF changes by at most 30\% over cosmic time, while SBR varies by a factor of three or more, making SBR the primary driver of STR evolution. The turnover in SBR at the critical mass naturally produces the observed STR turnover.

$\bullet$ In addition to this primary scale, we identify a secondary critical mass scale at $M_{\rm tot} \sim 10^{11} {\rm M_\odot}$ (or $M_* \sim 10^{9\text{--}9.5}{\rm M_\odot}$), characterized by the peak in the gas retention fraction $M_{\rm gas}/M_{\rm tot}$. Below this mass, abundant dynamically cool gas sustains efficient in-situ growth of STR. Above this mass, the hot gas fraction increases monotonically, leading to a gradual decline in in-situ star formation efficiency, as reflected in the turnover and monotonic decline in the rate at which in-situ star formation grows the STR.

Taken together, these results place several observed transitions in galaxy evolution within a unified physical framework. With an improved understanding of the physics associated with the critical mass scales, we plan to re-examine transitions in the collective physical properties of galaxies including star-formation activity, morphology, surface brightness, the mass-metallicity relation, size-mass relation, the slope of the red sequence, and the Faber-Jackson and Tully-Fisher relations.

\begin{acknowledgments}
PKM is supported by the KIAS Individual Grant (PG096702) at the Korea Institute for Advanced Study. CBP is supported by the KIAS Individual Grant PG016904 at the Korea Institute for Advanced Study (KIAS) and by the National Research Foundation of Korea (NRF) grant funded by the Korean government (MSIT, RS2024-00360385). JL also acknowledges the support of the NRF of Korea grant funded by the Korea government (MSIT, RS2022-NR068800). This work is partially supported by the grant GALBAR ANR-25-CE31-4684 and from the CNRS through the MITI interdisciplinary programs. This work is also supported by the Center for Advanced Computation at Korea Institute for Advanced Study.

\end{acknowledgments}


\software{AstroPy \citep{2013A&A...558A..33A,2018AJ....156..123A}, Matplotlib \citep{hunter2007}
          }



\appendix

\section{Evolution of energy injection rate from AGN} \label{AGNrate}

\begin{figure}
    \centering
    \includegraphics[width = 0.47\textwidth]{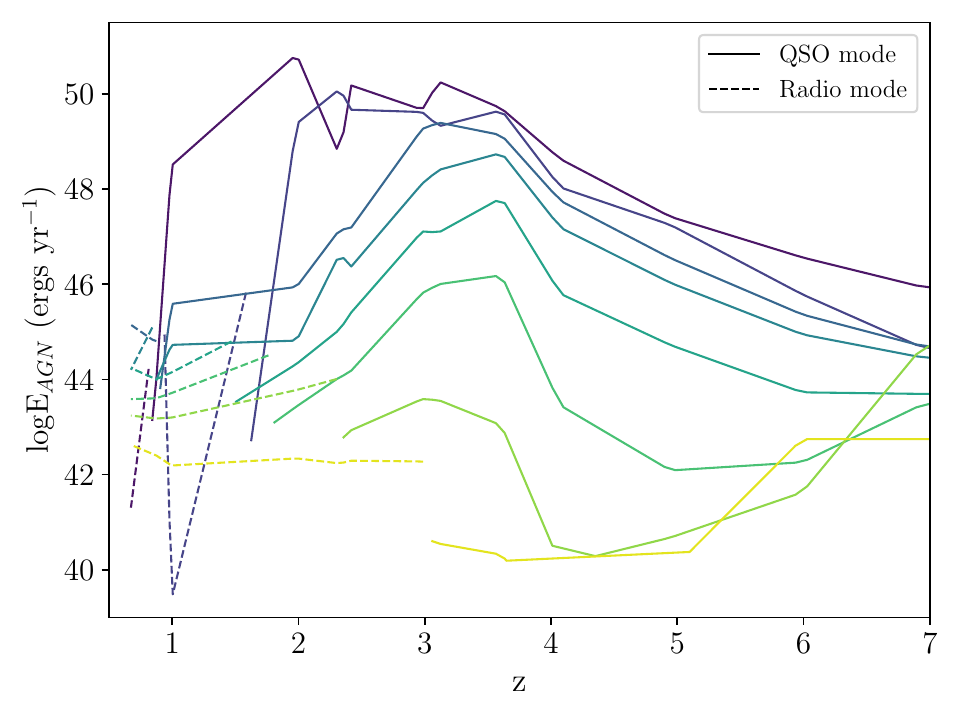}
    
    \caption{Median AGN injection rate as function of redshift. The lines denote the evolutionary trajectory of galaxies, color coded by their final total mass. The color coding scheme remains same as in Figure \ref{fig2}.}
    \label{A1}
\end{figure}

The AGN feedback in the HR5 simulation operates through two modes: a thermal-QSO mode and a kinetic jet-AGN mode, determined by the Eddington ratio, $\chi = \dot{M}_{\rm BH} / \dot{M}_{\rm Edd}$, where $\dot{M}_{\rm BH}$ is the black hole accretion rate and $\dot{M}_{\rm Edd}$ is the Eddington accretion limit. For $\chi > 0.01$, feedback occurs in the QSO-thermal mode, while for $\chi \leq 0.01$, it switches to the radio-jet mode. The energy injection rate is $\epsilon_r \epsilon_{f,h} \dot{M}_{\rm BHL} c^2$ for the QSO-thermal mode and $\epsilon{f,j} \dot{M}_{\rm BHL} c^2$ for the kinetic-jet mode, where $\dot{M}_{\rm BHL} = \dot{M}_{\rm BH} / (1 - \epsilon_r)$. The free parameters $\epsilon_r$, $\epsilon{f,h}$, and $\epsilon_{f,j}$ are calibrated to reproduce the observed cosmic star formation history, galaxy stellar mass function, and stellar mass–black hole mass relation. More details on the feedback prescriptions and parameters can be found in \cite{Lee2021}.

To compute the AGN energy injection rates for our sample galaxies, we query the relevant quantities at snapshots corresponding to redshifts $z = [0.625, 0.873, 0.995, 1.985, 2.336, 2.395, 2.98, 3.111, 3.608, 4.074, 4.946, 5.989, 6.96]$. We then calculate the median energy injection rates for both the thermal-QSO and jet-AGN modes, as well as the median value of $\chi$, for the progenitors of our sample galaxies at each snapshot. Based on the median $\chi$, we determine which feedback mode is effective at a given redshift. The median trend of AGN energy injection rate as a function of redshift is shown in Figure \ref{A1}. From the plot, we see that the AGN energy injection peaks roughly around $3 \lesssim z \lesssim 4$ for most galaxy progenitors, irrespective of their final mass. This suggests that AGN feedback is more closely tied to cosmic time than to a redshift-independent mass scale, and is therefore unlikely to be a primary driver of the two critical mass scales. However, AGN feedback may contribute by heating the gas halo and prevent its cooling, thereby suppressing star formation in massive galaxies, maintaining the decline of STR.

\section{Comparison of hot gas fraction with observations} \label{obs}

We analyze the hot gas fraction $f_{\rm hot} \equiv M_{\rm hot,gas}/M_{\rm tot}$ as a function of total mass for HR5 galaxies at $z=0.625$, computing the median and $16$th--$84$th percentile scatter in logarithmic mass bins of $0.3$ dex. We compare our simulation results against X-ray–derived hot gas fractions from \citet{Popesso2024} and \citet{Lyskova2023}, both at low redshift ($z<0.2$), as well as the empirical $f_{\rm hot}$–$M_{200}$ scaling relation from \citet{Popesso2024}. We note that our hot gas definition adopts a temperature threshold of $T \geq 3\times10^4$ K, which is lower than the temperatures primarily traced by X-ray observations, and that the observational results are at lower redshifts than the HR5 results. Despite these differences, we find good overall agreement between the simulation and observational trends.

\begin{figure}
    \centering
    \includegraphics[width = 0.495\textwidth]{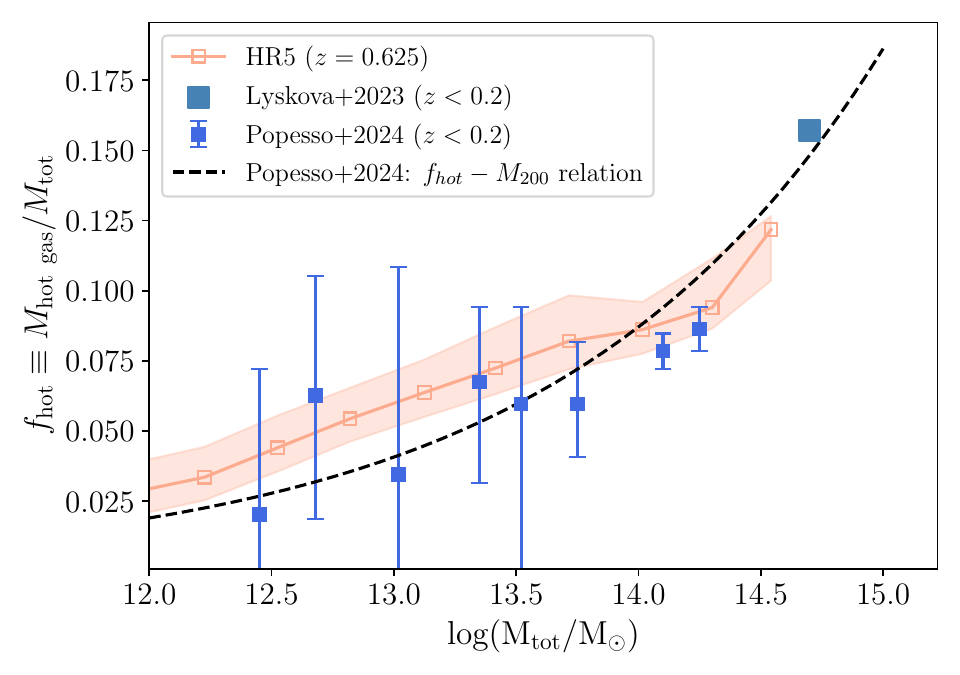}
    
    \caption{Hot gas fraction ($f_{\rm hot} \equiv M_{\rm hot\,gas}/M_{\rm tot}$) as a function of total mass for galaxies from the HR5 simulation at $z=0.625$ (red line). The shaded region represents the $16$th--$84$th percentile scatter in each bin. Observational data from \citet{Lyskova2023} ($z<0.2$) and \citet{Popesso2024} ($z<0.2$) are shown as blue symbols. The dashed black line shows the empirical $f_{\rm hot}$--$M_{200}$ scaling relation from \citet{Popesso2024}.}
    \label{A0}
\end{figure}

\bibliography{sample631}{}

\begin{thebibliography}{}
\expandafter\ifx\csname natexlab\endcsname\relax\def\natexlab#1{#1}\fi
\providecommand{\url}[1]{\href{#1}{#1}}
\providecommand{\dodoi}[1]{doi:~\href{http://doi.org/#1}{\nolinkurl{#1}}}
\providecommand{\doeprint}[1]{\href{http://ascl.net/#1}{\nolinkurl{http://ascl.net/#1}}}
\providecommand{\doarXiv}[1]{\href{https://arxiv.org/abs/#1}{\nolinkurl{https://arxiv.org/abs/#1}}}

\bibitem[{ {Astropy Collaboration} {et~al.}(2013){Astropy Collaboration}, {Robitaille}, {Tollerud}, {Greenfield}, {Droettboom}, {Bray}, {Aldcroft}, {Davis}, {Ginsburg}, {Price-Whelan}, {Kerzendorf}, {Conley}, {Crighton}, {Barbary}, {Muna}, {Ferguson}, {Grollier}, {Parikh}, {Nair}, {Unther}, {Deil}, {Woillez}, {Conseil}, {Kramer}, {Turner}, {Singer}, {Fox}, {Weaver}, {Zabalza}, {Edwards}, {Azalee Bostroem}, {Burke}, {Casey}, {Crawford}, {Dencheva}, {Ely}, {Jenness}, {Labrie}, {Lim}, {Pierfederici}, {Pontzen}, {Ptak}, {Refsdal}, {Servillat}, \& {Streicher}}]{2013A&A...558A..33A}
{Astropy Collaboration}, {Robitaille}, T.~P., {Tollerud}, E.~J., {et~al.} 2013, \bibinfo{title}{{Astropy: A community Python package for astronomy},} \aap, 558, A33, \dodoi{10.1051/0004-6361/201322068}

\bibitem[{ {Astropy Collaboration} {et~al.}(2018){Astropy Collaboration}, {Price-Whelan}, {Sip{\H{o}}cz}, {G{\"u}nther}, {Lim}, {Crawford}, {Conseil}, {Shupe}, {Craig}, {Dencheva}, {Ginsburg}, {VanderPlas}, {Bradley}, {P{\'e}rez-Su{\'a}rez}, {de Val-Borro}, {Aldcroft}, {Cruz}, {Robitaille}, {Tollerud}, {Ardelean}, {Babej}, {Bach}, {Bachetti}, {Bakanov}, {Bamford}, {Barentsen}, {Barmby}, {Baumbach}, {Berry}, {Biscani}, {Boquien}, {Bostroem}, {Bouma}, {Brammer}, {Bray}, {Breytenbach}, {Buddelmeijer}, {Burke}, {Calderone}, {Cano Rodr{\'\i}guez}, {Cara}, {Cardoso}, {Cheedella}, {Copin}, {Corrales}, {Crichton}, {D'Avella}, {Deil}, {Depagne}, {Dietrich}, {Donath}, {Droettboom}, {Earl}, {Erben}, {Fabbro}, {Ferreira}, {Finethy}, {Fox}, {Garrison}, {Gibbons}, {Goldstein}, {Gommers}, {Greco}, {Greenfield}, {Groener}, {Grollier}, {Hagen}, {Hirst}, {Homeier}, {Horton}, {Hosseinzadeh}, {Hu}, {Hunkeler}, {Ivezi{\'c}}, {Jain}, {Jenness}, {Kanarek}, {Kendrew}, {Kern}, {Kerzendorf}, {Khvalko}, {King}, {Kirkby}, {Kulkarni},
  {Kumar}, {Lee}, {Lenz}, {Littlefair}, {Ma}, {Macleod}, {Mastropietro}, {McCully}, {Montagnac}, {Morris}, {Mueller}, {Mumford}, {Muna}, {Murphy}, {Nelson}, {Nguyen}, {Ninan}, {N{\"o}the}, {Ogaz}, {Oh}, {Parejko}, {Parley}, {Pascual}, {Patil}, {Patil}, {Plunkett}, {Prochaska}, {Rastogi}, {Reddy Janga}, {Sabater}, {Sakurikar}, {Seifert}, {Sherbert}, {Sherwood-Taylor}, {Shih}, {Sick}, {Silbiger}, {Singanamalla}, {Singer}, {Sladen}, {Sooley}, {Sornarajah}, {Streicher}, {Teuben}, {Thomas}, {Tremblay}, {Turner}, {Terr{\'o}n}, {van Kerkwijk}, {de la Vega}, {Watkins}, {Weaver}, {Whitmore}, {Woillez}, {Zabalza}, \& {Astropy Contributors}}]{2018AJ....156..123A}
{Astropy Collaboration}, {Price-Whelan}, A.~M., {Sip{\H{o}}cz}, B.~M., {et~al.} 2018, \bibinfo{title}{{The Astropy Project: Building an Open-science Project and Status of the v2.0 Core Package},} \aj, 156, 123, \dodoi{10.3847/1538-3881/aabc4f}

\bibitem[{I.~K. {Baldry} {et~al.}(2004){Baldry}, {Glazebrook}, {Brinkmann}, {Ivezi{\'c}}, {Lupton}, {Nichol}, \& {Szalay}}]{Baldry2004}
{Baldry}, I.~K., {Glazebrook}, K., {Brinkmann}, J., {et~al.} 2004, \bibinfo{title}{{Quantifying the Bimodal Color-Magnitude Distribution of Galaxies},} \apj, 600, 681, \dodoi{10.1086/380092}

\bibitem[{P.~S. {Behroozi} {et~al.}(2013){Behroozi}, {Wechsler}, \& {Conroy}}]{Behroozi2013}
{Behroozi}, P.~S., {Wechsler}, R.~H., \& {Conroy}, C. 2013, \bibinfo{title}{{The Average Star Formation Histories of Galaxies in Dark Matter Halos from z = 0-8},} \apj, 770, 57, \dodoi{10.1088/0004-637X/770/1/57}

\bibitem[{G.~A. {Blanc} {et~al.}(2019){Blanc}, {Lu}, {Benson}, {Katsianis}, \& {Barraza}}]{Blanc2019}
{Blanc}, G.~A., {Lu}, Y., {Benson}, A., {Katsianis}, A., \& {Barraza}, M. 2019, \bibinfo{title}{{A Characteristic Mass Scale in the Mass-Metallicity Relation of Galaxies},} \apj, 877, 6, \dodoi{10.3847/1538-4357/ab16ec}

\bibitem[{C.~M. {Booth} \& J. {Schaye}(2010){Booth} \& {Schaye}}]{Booth2010}
{Booth}, C.~M., \& {Schaye}, J. 2010, \bibinfo{title}{{Dark matter haloes determine the masses of supermassive black holes},} \mnras, 405, L1, \dodoi{10.1111/j.1745-3933.2010.00832.x}

\bibitem[{C.~M. {Booth} \& J. {Schaye}(2011){Booth} \& {Schaye}}]{Booth2011}
{Booth}, C.~M., \& {Schaye}, J. 2011, \bibinfo{title}{{Towards an understanding of the evolution of the scaling relations for supermassive black holes},} \mnras, 413, 1158, \dodoi{10.1111/j.1365-2966.2011.18203.x}

\bibitem[{P. {Boubel} {et~al.}(2024){Boubel}, {Colless}, {Said}, \& {Staveley-Smith}}]{Boubel2024}
{Boubel}, P., {Colless}, M., {Said}, K., \& {Staveley-Smith}, L. 2024, \bibinfo{title}{{Large-scale motions and growth rate from forward-modelling Tully-Fisher peculiar velocities},} \mnras, 531, 84, \dodoi{10.1093/mnras/stae1122}

\bibitem[{Y.-Y. {Choi} {et~al.}(2007){Choi}, {Park}, \& {Vogeley}}]{Choi2007}
{Choi}, Y.-Y., {Park}, C., \& {Vogeley}, M.~S. 2007, \bibinfo{title}{{Internal and Collective Properties of Galaxies in the Sloan Digital Sky Survey},} \apj, 658, 884, \dodoi{10.1086/511060}

\bibitem[{C.~A. {Correa} {et~al.}(2018){Correa}, {Schaye}, {Wyithe}, {Duffy}, {Theuns}, {Crain}, \& {Bower}}]{Correa2018}
{Correa}, C.~A., {Schaye}, J., {Wyithe}, J. S.~B., {et~al.} 2018, \bibinfo{title}{{The formation of hot gaseous haloes around galaxies},} \mnras, 473, 538, \dodoi{10.1093/mnras/stx2332}

\bibitem[{M. {Curti} {et~al.}(2024){Curti}, {Maiolino}, {Curtis-Lake}, {Chevallard}, {Carniani}, {D'Eugenio}, {Looser}, {Scholtz}, {Charlot}, {Cameron}, {{\"U}bler}, {Witstok}, {Boyett}, {Laseter}, {Sandles}, {Arribas}, {Bunker}, {Giardino}, {Maseda}, {Rawle}, {Rodr{\'\i}guez Del Pino}, {Smit}, {Willott}, {Eisenstein}, {Hausen}, {Johnson}, {Rieke}, {Robertson}, {Tacchella}, {Williams}, {Willmer}, {Baker}, {Bhatawdekar}, {Egami}, {Helton}, {Ji}, {Kumari}, {Perna}, {Shivaei}, \& {Sun}}]{Curtis2024}
{Curti}, M., {Maiolino}, R., {Curtis-Lake}, E., {et~al.} 2024, \bibinfo{title}{{JADES: Insights into the low-mass end of the mass-metallicity-SFR relation at 3 < z < 10 from deep JWST/NIRSpec spectroscopy},} \aap, 684, A75, \dodoi{10.1051/0004-6361/202346698}

\bibitem[{L.~J.~M. {Davies} {et~al.}(2019){Davies}, {Robotham}, {Lagos}, {Driver}, {Stevens}, {Bah{\'e}}, {Alpaslan}, {Bremer}, {Brown}, {Brough}, {Bland-Hawthorn}, {Cortese}, {Elahi}, {Grootes}, {Holwerda}, {Ludlow}, {McGee}, {Owers}, \& {Phillipps}}]{Davies2019}
{Davies}, L.~J.~M., {Robotham}, A.~S.~G., {Lagos}, C. d.~P., {et~al.} 2019, \bibinfo{title}{{Galaxy and Mass Assembly (GAMA): environmental quenching of centrals and satellites in groups},} \mnras, 483, 5444, \dodoi{10.1093/mnras/sty3393}

\bibitem[{A. Dekel \& Y. Birnboim(2006)Dekel \& Birnboim}]{Dekel_2006}
Dekel, A., \& Birnboim, Y. 2006, \bibinfo{title}{Galaxy bimodality due to cold flows and shock heating,} Monthly Notices of the Royal Astronomical Society, 368, 2–20, \dodoi{10.1111/j.1365-2966.2006.10145.x}

\bibitem[{A. Dekel {et~al.}(2019)Dekel, Lapiner, \& Dubois}]{Dekel_2019}
Dekel, A., Lapiner, S., \& Dubois, Y. 2019, \bibinfo{title}{Origin of the Golden Mass of Galaxies and Black Holes,} \doarXiv{1904.08431}

\bibitem[{Y. {Dubois} {et~al.}(2016){Dubois}, {Peirani}, {Pichon}, {Devriendt}, {Gavazzi}, {Welker}, \& {Volonteri}}]{Dubois2016}
{Dubois}, Y., {Peirani}, S., {Pichon}, C., {et~al.} 2016, \bibinfo{title}{{The HORIZON-AGN simulation: morphological diversity of galaxies promoted by AGN feedback},} \mnras, 463, 3948, \dodoi{10.1093/mnras/stw2265}

\bibitem[{Y. {Dubois} {et~al.}(2015){Dubois}, {Volonteri}, {Silk}, {Devriendt}, {Slyz}, \& {Teyssier}}]{Dubois2015}
{Dubois}, Y., {Volonteri}, M., {Silk}, J., {et~al.} 2015, \bibinfo{title}{{Black hole evolution - I. Supernova-regulated black hole growth},} \mnras, 452, 1502, \dodoi{10.1093/mnras/stv1416}

\bibitem[{A. Enia {et~al.}(2022)Enia, Talia, Pozzi, Cimatti, Delvecchio, Zamorani, D’Amato, Bisigello, Gruppioni, Rodighiero, Calura, Dallacasa, Giulietti, Barchiesi, Behiri, \& Romano}]{Enia_2022}
Enia, A., Talia, M., Pozzi, F., {et~al.} 2022, \bibinfo{title}{A New Estimate of the Cosmic Star Formation Density from a Radio-selected Sample, and the Contribution of H-dark Galaxies at z ≥ 3,} The Astrophysical Journal, 927, 204, \dodoi{10.3847/1538-4357/ac51ca}

\bibitem[{C. {Gruppioni} {et~al.}(2013){Gruppioni}, {Pozzi}, {Rodighiero}, {Delvecchio}, {Berta}, {Pozzetti}, {Zamorani}, {Andreani}, {Cimatti}, {Ilbert}, {Le Floc'h}, {Lutz}, {Magnelli}, {Marchetti}, {Monaco}, {Nordon}, {Oliver}, {Popesso}, {Riguccini}, {Roseboom}, {Rosario}, {Sargent}, {Vaccari}, {Altieri}, {Aussel}, {Bongiovanni}, {Cepa}, {Daddi}, {Dom{\'\i}nguez-S{\'a}nchez}, {Elbaz}, {F{\"o}rster Schreiber}, {Genzel}, {Iribarrem}, {Magliocchetti}, {Maiolino}, {Poglitsch}, {P{\'e}rez Garc{\'\i}a}, {Sanchez-Portal}, {Sturm}, {Tacconi}, {Valtchanov}, {Amblard}, {Arumugam}, {Bethermin}, {Bock}, {Boselli}, {Buat}, {Burgarella}, {Castro-Rodr{\'\i}guez}, {Cava}, {Chanial}, {Clements}, {Conley}, {Cooray}, {Dowell}, {Dwek}, {Eales}, {Franceschini}, {Glenn}, {Griffin}, {Hatziminaoglou}, {Ibar}, {Isaak}, {Ivison}, {Lagache}, {Levenson}, {Lu}, {Madden}, {Maffei}, {Mainetti}, {Nguyen}, {O'Halloran}, {Page}, {Panuzzo}, {Papageorgiou}, {Pearson}, {P{\'e}rez-Fournon}, {Pohlen}, {Rigopoulou}, {Rowan-Robinson}, {Schulz},
  {Scott}, {Seymour}, {Shupe}, {Smith}, {Stevens}, {Symeonidis}, {Trichas}, {Tugwell}, {Vigroux}, {Wang}, {Wright}, {Xu}, {Zemcov}, {Bardelli}, {Carollo}, {Contini}, {Le F{\'e}vre}, {Lilly}, {Mainieri}, {Renzini}, {Scodeggio}, \& {Zucca}}]{Grupponi2013}
{Gruppioni}, C., {Pozzi}, F., {Rodighiero}, G., {et~al.} 2013, \bibinfo{title}{{The Herschel PEP/HerMES luminosity function - I. Probing the evolution of PACS selected Galaxies to z ≃ 4},} \mnras, 432, 23, \dodoi{10.1093/mnras/stt308}

\bibitem[{C. {Gruppioni} {et~al.}(2020){Gruppioni}, {B{\'e}thermin}, {Loiacono}, {Le F{\`e}vre}, {Capak}, {Cassata}, {Faisst}, {Schaerer}, {Silverman}, {Yan}, {Bardelli}, {Boquien}, {Carraro}, {Cimatti}, {Dessauges-Zavadsky}, {Ginolfi}, {Fujimoto}, {Hathi}, {Jones}, {Khusanova}, {Koekemoer}, {Lagache}, {Lemaux}, {Oesch}, {Pozzi}, {Riechers}, {Rodighiero}, {Romano}, {Talia}, {Vallini}, {Vergani}, {Zamorani}, \& {Zucca}}]{Gruppioni2020}
{Gruppioni}, C., {B{\'e}thermin}, M., {Loiacono}, F., {et~al.} 2020, \bibinfo{title}{{The ALPINE-ALMA [CII] survey. The nature, luminosity function, and star formation history of dusty galaxies up to z ≃ 6},} \aap, 643, A8, \dodoi{10.1051/0004-6361/202038487}

\bibitem[{M. {Habouzit} {et~al.}(2017){Habouzit}, {Volonteri}, \& {Dubois}}]{Habouzit2017}
{Habouzit}, M., {Volonteri}, M., \& {Dubois}, Y. 2017, \bibinfo{title}{{Blossoms from black hole seeds: properties and early growth regulated by supernova feedback},} \mnras, 468, 3935, \dodoi{10.1093/mnras/stx666}

\bibitem[{O. {Hahn} \& T. {Abel}(2011){Hahn} \& {Abel}}]{Hahn2011}
{Hahn}, O., \& {Abel}, T. 2011, \bibinfo{title}{{Multi-scale initial conditions for cosmological simulations},} \mnras, 415, 2101, \dodoi{10.1111/j.1365-2966.2011.18820.x}

\bibitem[{S.~E. {Hong} {et~al.}(2024){Hong}, {Park}, {Mishra}, {Kim}, {Gibson}, {Kim}, {Few}, {Pichon}, {Shin}, \& {Lee}}]{Hong2024}
{Hong}, S.~E., {Park}, C., {Mishra}, P.~K., {et~al.} 2024, \bibinfo{title}{{Emergence of the Galaxy Morphology{\textendash}Star Formation Activity{\textendash}Clustercentric Radius Relations in Galaxy Clusters},} \apj, 977, 183, \dodoi{10.3847/1538-4357/ad9276}

\bibitem[{A.~M. {Hopkins}(2004){Hopkins}}]{Hopkins2004}
{Hopkins}, A.~M. 2004, \bibinfo{title}{{On the Evolution of Star-forming Galaxies},} \apj, 615, 209, \dodoi{10.1086/424032}

\bibitem[{S. {Huang} {et~al.}(2012){Huang}, {Haynes}, {Giovanelli}, \& {Brinchmann}}]{Huang2012}
{Huang}, S., {Haynes}, M.~P., {Giovanelli}, R., \& {Brinchmann}, J. 2012, \bibinfo{title}{{The Arecibo Legacy Fast ALFA Survey: The Galaxy Population Detected by ALFALFA},} \apj, 756, 113, \dodoi{10.1088/0004-637X/756/2/113}

\bibitem[{J.~D. Hunter(2007)Hunter}]{hunter2007}
Hunter, J.~D. 2007, \bibinfo{title}{Matplotlib: A 2D graphics environment,} Computing in Science \& Engineering, 9, 90, \dodoi{10.1109/MCSE.2007.55}

\bibitem[{D. {Jeong} {et~al.}(2025){Jeong}, {Hwang}, {Chung}, \& {Yoon}}]{Jeong2025}
{Jeong}, D., {Hwang}, H.~S., {Chung}, H., \& {Yoon}, Y. 2025, \bibinfo{title}{{Diverse Rotation Curves of Galaxies in a Simulated Universe: The Observed Dependence on Stellar Mass and Morphology Reproduced},} \apj, 982, 11, \dodoi{10.3847/1538-4357/adb1be}

\bibitem[{G. {Kauffmann} {et~al.}(2003){Kauffmann}, {Heckman}, {White}, {Charlot}, {Tremonti}, {Brinchmann}, {Bruzual}, {Peng}, {Seibert}, {Bernardi}, {Blanton}, {Brinkmann}, {Castander}, {Cs{\'a}bai}, {Fukugita}, {Ivezic}, {Munn}, {Nichol}, {Padmanabhan}, {Thakar}, {Weinberg}, \& {York}}]{Kauffman2004}
{Kauffmann}, G., {Heckman}, T.~M., {White}, S. D.~M., {et~al.} 2003, \bibinfo{title}{{Stellar masses and star formation histories for {}10$^{5}$ galaxies from the Sloan Digital Sky Survey},} \mnras, 341, 33, \dodoi{10.1046/j.1365-8711.2003.06291.x}

\bibitem[{L. {Kawinwanichakij} {et~al.}(2021){Kawinwanichakij}, {Silverman}, {Ding}, {George}, {Damjanov}, {Sawicki}, {Tanaka}, {Taranu}, {Birrer}, {Huang}, {Li}, {Onodera}, {Shibuya}, \& {Yasuda}}]{Nancy2021}
{Kawinwanichakij}, L., {Silverman}, J.~D., {Ding}, X., {et~al.} 2021, \bibinfo{title}{{Hyper Suprime-Cam Subaru Strategic Program: A Mass-dependent Slope of the Galaxy Size-Mass Relation at z < 1},} \apj, 921, 38, \dodoi{10.3847/1538-4357/ac1f21}

\bibitem[{L.~S. {Kelvin} {et~al.}(2014){Kelvin}, {Driver}, {Robotham}, {Taylor}, {Graham}, {Alpaslan}, {Baldry}, {Bamford}, {Bauer}, {Bland-Hawthorn}, {Brown}, {Colless}, {Conselice}, {Holwerda}, {Hopkins}, {Lara-L{\'o}pez}, {Liske}, {L{\'o}pez-S{\'a}nchez}, {Loveday}, {Norberg}, {Phillipps}, {Popescu}, {Prescott}, {Sansom}, \& {Tuffs}}]{Kelvin2014}
{Kelvin}, L.~S., {Driver}, S.~P., {Robotham}, A. S.~G., {et~al.} 2014, \bibinfo{title}{{Galaxy And Mass Assembly (GAMA): stellar mass functions by Hubble type},} \mnras, 444, 1647, \dodoi{10.1093/mnras/stu1507}

\bibitem[{M.~G. {Kendall} \& A. {Stuart}(1977){Kendall} \& {Stuart}}]{KendallStuart1977}
{Kendall}, M.~G., \& {Stuart}, A. 1977, The Advanced Theory of Statistics, Vol.~1: Distribution Theory, 4th edn. (London: Charles Griffin \& Company)

\bibitem[{D. {Kere{\v{s}}} {et~al.}(2005){Kere{\v{s}}}, {Katz}, {Weinberg}, \& {Dav{\'e}}}]{Keres2005}
{Kere{\v{s}}}, D., {Katz}, N., {Weinberg}, D.~H., \& {Dav{\'e}}, R. 2005, \bibinfo{title}{{How do galaxies get their gas?},} \mnras, 363, 2, \dodoi{10.1111/j.1365-2966.2005.09451.x}

\bibitem[{J. Kim \& C. Park(2006)Kim \& Park}]{Kim2006}
Kim, J., \& Park, C. 2006, \bibinfo{title}{A New Halo‐finding Method forN‐Body Simulations,} The Astrophysical Journal, 639, 600–616, \dodoi{10.1086/499761}

\bibitem[{J. {Kim} {et~al.}(2023){Kim}, {Lee}, {Laigle}, {Dubois}, {Kim}, {Park}, {Pichon}, {Gibson}, {Few}, {Shin}, \& {Snaith}}]{Kim2023}
{Kim}, J., {Lee}, J., {Laigle}, C., {et~al.} 2023, \bibinfo{title}{{Low-surface-brightness Galaxies are Missing in the Observed Stellar Mass Function},} \apj, 951, 137, \dodoi{10.3847/1538-4357/acd251}

\bibitem[{A.~V. {Kravtsov} {et~al.}(2018){Kravtsov}, {Vikhlinin}, \& {Meshcheryakov}}]{Kravtsov2018}
{Kravtsov}, A.~V., {Vikhlinin}, A.~A., \& {Meshcheryakov}, A.~V. 2018, \bibinfo{title}{{Stellar Mass{\textemdash}Halo Mass Relation and Star Formation Efficiency in High-Mass Halos},} Astronomy Letters, 44, 8, \dodoi{10.1134/S1063773717120015}

\bibitem[{S. {Lapiner} {et~al.}(2021){Lapiner}, {Dekel}, \& {Dubois}}]{Lapiner2021}
{Lapiner}, S., {Dekel}, A., \& {Dubois}, Y. 2021, \bibinfo{title}{{Compaction-driven black hole growth},} \mnras, 505, 172, \dodoi{10.1093/mnras/stab1205}

\bibitem[{S. {Lapiner} {et~al.}(2023){Lapiner}, {Dekel}, {Freundlich}, {Ginzburg}, {Jiang}, {Kretschmer}, {Tacchella}, {Ceverino}, \& {Primack}}]{Lapiner2023}
{Lapiner}, S., {Dekel}, A., {Freundlich}, J., {et~al.} 2023, \bibinfo{title}{{Wet compaction to a blue nugget: a critical phase in galaxy evolution},} \mnras, 522, 4515, \dodoi{10.1093/mnras/stad1263}

\bibitem[{M.~A. Lara-López {et~al.}(2013)Lara-López, Hopkins, López-Sánchez, Brough, Gunawardhana, Colless, Robotham, Bauer, Bland-Hawthorn, Cluver, Driver, Foster, Kelvin, Liske, Loveday, Owers, Ponman, Sharp, Steele, Taylor, \& Thomas}]{Lopez2013}
Lara-López, M.~A., Hopkins, A.~M., López-Sánchez, A.~R., {et~al.} 2013, \bibinfo{title}{Galaxy And Mass Assembly (GAMA): a deeper view of the mass, metallicity and SFR relationships,} Monthly Notices of the Royal Astronomical Society, 434, 451–470, \dodoi{10.1093/mnras/stt1031}

\bibitem[{A. {Le Bail} {et~al.}(2024){Le Bail}, {Daddi}, {Elbaz}, {Dickinson}, {Giavalisco}, {Magnelli}, {G{\'o}mez-Guijarro}, {Kalita}, {Koekemoer}, {Holwerda}, {Bournaud}, {de la Vega}, {Calabr{\`o}}, {Dekel}, {Cheng}, {Bisigello}, {Franco}, {Costantin}, {Lucas}, {P{\'e}rez-Gonz{\'a}lez}, {Lu}, {Wilkins}, {Arrabal Haro}, {Bagley}, {Finkelstein}, {Kartaltepe}, {Papovich}, {Pirzkal}, \& {Yung}}]{Bail2024}
{Le Bail}, A., {Daddi}, E., {Elbaz}, D., {et~al.} 2024, \bibinfo{title}{{JWST/CEERS sheds light on dusty star-forming galaxies: Forming bulges, lopsidedness, and outside-in quenching at cosmic noon},} \aap, 688, A53, \dodoi{10.1051/0004-6361/202347465}

\bibitem[{J. {Lee} {et~al.}(2021){Lee}, {Shin}, {Snaith}, {Kim}, {Few}, {Devriendt}, {Dubois}, {Cox}, {Hong}, {Kwon}, {Park}, {Pichon}, {Kim}, {Gibson}, \& {Park}}]{Lee2021}
{Lee}, J., {Shin}, J., {Snaith}, O.~N., {et~al.} 2021, \bibinfo{title}{{The Horizon Run 5 Cosmological Hydrodynamical Simulation: Probing Galaxy Formation from Kilo- to Gigaparsec Scales},} \apj, 908, 11, \dodoi{10.3847/1538-4357/abd08b}

\bibitem[{J.~H. {Lee} {et~al.}(2024){Lee}, {Park}, {Hwang}, \& {Kwon}}]{JHLee2024}
{Lee}, J.~H., {Park}, C., {Hwang}, H.~S., \& {Kwon}, M. 2024, \bibinfo{title}{{Morphology of Galaxies in JWST Fields: Initial Distribution and Evolution of Galaxy Morphology},} \apj, 966, 113, \dodoi{10.3847/1538-4357/ad3448}

\bibitem[{A. {Lewis} {et~al.}(2000){Lewis}, {Challinor}, \& {Lasenby}}]{Lewis2000}
{Lewis}, A., {Challinor}, A., \& {Lasenby}, A. 2000, \bibinfo{title}{{Efficient Computation of Cosmic Microwave Background Anisotropies in Closed Friedmann-Robertson-Walker Models},} \apj, 538, 473, \dodoi{10.1086/309179}

\bibitem[{L. {Lucie-Smith} {et~al.}(2025){Lucie-Smith}, {Peiris}, {Pontzen}, {Halder}, {Schaye}, {Schaller}, {Helly}, {McGibbon}, \& {Elbers}}]{LucieSmith2025}
{Lucie-Smith}, L., {Peiris}, H.~V., {Pontzen}, A., {et~al.} 2025, \bibinfo{title}{{Cosmological feedback from a halo assembly perspective},} \prd, 112, 063541, \dodoi{10.1103/vh8n-9cr2}

\bibitem[{N. {Lyskova} {et~al.}(2023){Lyskova}, {Churazov}, {Khabibullin}, {Burenin}, {Starobinsky}, \& {Sunyaev}}]{Lyskova2023}
{Lyskova}, N., {Churazov}, E., {Khabibullin}, I.~I., {et~al.} 2023, \bibinfo{title}{{X-ray surface brightness and gas density profiles of galaxy clusters up to 3 {\texttimes} R$_{500c}$ with SRG/eROSITA},} \mnras, 525, 898, \dodoi{10.1093/mnras/stad2305}

\bibitem[{P. {Madau} \& M. {Dickinson}(2014){Madau} \& {Dickinson}}]{Madau_Dickinson2014}
{Madau}, P., \& {Dickinson}, M. 2014, \bibinfo{title}{{Cosmic Star-Formation History},} \araa, 52, 415, \dodoi{10.1146/annurev-astro-081811-125615}

\bibitem[{P. {Madau} {et~al.}(1998){Madau}, {Pozzetti}, \& {Dickinson}}]{Madau1998}
{Madau}, P., {Pozzetti}, L., \& {Dickinson}, M. 1998, \bibinfo{title}{{The Star Formation History of Field Galaxies},} \apj, 498, 106, \dodoi{10.1086/305523}

\bibitem[{B. {Magnelli} {et~al.}(2023){Magnelli}, {G{\'o}mez-Guijarro}, {Elbaz}, {Daddi}, {Papovich}, {Shen}, {Arrabal Haro}, {Bagley}, {Bell}, {Buat}, {Costantin}, {Dickinson}, {Finkelstein}, {Gardner}, {Jim{\'e}nez-Andrade}, {Kartaltepe}, {Koekemoer}, {Lyu}, {P{\'e}rez-Gonz{\'a}lez}, {Pirzkal}, {Tacchella}, {de la Vega}, {Wuyts}, {Yang}, {Yung}, \& {Zavala}}]{Magnelli2023}
{Magnelli}, B., {G{\'o}mez-Guijarro}, C., {Elbaz}, D., {et~al.} 2023, \bibinfo{title}{{CEERS: MIRI deciphers the spatial distribution of dust-obscured star formation in galaxies at 0.1 < z < 2.5},} \aap, 678, A83, \dodoi{10.1051/0004-6361/202347052}

\bibitem[{R.~M. {M{\'e}rida} {et~al.}(2025){M{\'e}rida}, {Sawicki}, {Iyer}, {Noirot}, {Willott}, {Brada{\v{c}}}, {Desprez}, {Martis}, {Muzzin}, {Rihtar{\v{s}}i{\v{c}}}, {Sarrouh}, {Favaro}, {Gaspar}, {Harshan}, \& {Jude{\v{z}}}}]{Merida2025}
{M{\'e}rida}, R.~M., {Sawicki}, M., {Iyer}, K.~G., {et~al.} 2025, \bibinfo{title}{{Probing the Star Formation Main Sequence down to 10$^{7} M_\odot$ at $1 < z < 9$},} arXiv e-prints, arXiv:2509.22871, \dodoi{10.48550/arXiv.2509.22871}

\bibitem[{P.~K. {Mishra} {et~al.}(2023){Mishra}, {Rana}, \& {More}}]{Mishra2023}
{Mishra}, P.~K., {Rana}, D., \& {More}, S. 2023, \bibinfo{title}{{Stellar mass dependence of galaxy size-dark matter halo radius relation probed by Subaru-HSC survey weak lensing measurements},} \mnras, 526, 2403, \dodoi{10.1093/mnras/stad2914}

\bibitem[{L. Mowla {et~al.}(2019)Mowla, Wel, Dokkum, \& Miller}]{Mowla_2019}
Mowla, L., Wel, A. v.~d., Dokkum, P.~v., \& Miller, T.~B. 2019, \bibinfo{title}{A Mass-dependent Slope of the Galaxy Size–Mass Relation out to z ∼ 3: Further Evidence for a Direct Relation between Median Galaxy Size and Median Halo Mass,} The Astrophysical Journal Letters, 872, L13, \dodoi{10.3847/2041-8213/ab0379}

\bibitem[{M. {Novak} {et~al.}(2017){Novak}, {Smol{\v{c}}i{\'c}}, {Delhaize}, {Delvecchio}, {Zamorani}, {Baran}, {Bondi}, {Capak}, {Carilli}, {Ciliegi}, {Civano}, {Ilbert}, {Karim}, {Laigle}, {Le F{\`e}vre}, {Marchesi}, {McCracken}, {Miettinen}, {Salvato}, {Sargent}, {Schinnerer}, \& {Tasca}}]{Novac2017}
{Novak}, M., {Smol{\v{c}}i{\'c}}, V., {Delhaize}, J., {et~al.} 2017, \bibinfo{title}{{The VLA-COSMOS 3 GHz Large Project: Cosmic star formation history since z 5},} \aap, 602, A5, \dodoi{10.1051/0004-6361/201629436}

\bibitem[{P. {Ocvirk} {et~al.}(2008){Ocvirk}, {Pichon}, \& {Teyssier}}]{Ocvirk2008}
{Ocvirk}, P., {Pichon}, C., \& {Teyssier}, R. 2008, \bibinfo{title}{{Bimodal gas accretion in the Horizon-MareNostrum galaxy formation simulation},} \mnras, 390, 1326, \dodoi{10.1111/j.1365-2966.2008.13763.x}

\bibitem[{L. {Paquereau} {et~al.}(2025){Paquereau}, {Laigle}, {McCracken}, {Shuntov}, {Ilbert}, {Akins}, {Allen}, {Arango-Togo}, {Berman}, {B{\'e}thermin}, {Casey}, {McCleary}, {Dubois}, {Drakos}, {Faisst}, {Franco}, {Harish}, {Jespersen}, {Kartaltepe}, {Koekemoer}, {Kokorev}, {Lambrides}, {Larson}, {Liu}, {Le Borgne}, {Lewis}, {McKinney}, {Mercier}, {Rhodes}, {Robertson}, {Toft}, {Trebitsch}, {Tresse}, \& {Weaver}}]{Paquereau2025}
{Paquereau}, L., {Laigle}, C., {McCracken}, H.~J., {et~al.} 2025, \bibinfo{title}{{Tracing the galaxy-halo connection with galaxy clustering in COSMOS-Web from z = 0.1 to z {\ensuremath{\sim}} 12},} \aap, 702, A163, \dodoi{10.1051/0004-6361/202553828}

\bibitem[{C. Park {et~al.}(2007)Park, Choi, Vogeley, Gott~III, \& Blanton}]{Park2007}
Park, C., Choi, Y., Vogeley, M.~S., Gott~III, J.~R., \& Blanton, M.~R. 2007, \bibinfo{title}{Environmental Dependence of Properties of Galaxies in the Sloan Digital Sky Survey,} The Astrophysical Journal, 658, 898–916, \dodoi{10.1086/511059}

\bibitem[{C. {Park} \& Y.-Y. {Choi}(2009){Park} \& {Choi}}]{Park2009}
{Park}, C., \& {Choi}, Y.-Y. 2009, \bibinfo{title}{{Combined Effects of Galaxy Interactions and Large-Scale Environment on Galaxy Properties},} \apj, 691, 1828, \dodoi{10.1088/0004-637X/691/2/1828}

\bibitem[{C. {Park} {et~al.}(2008){Park}, {Gott}, \& {Choi}}]{Park2008}
{Park}, C., {Gott}, III, J.~R., \& {Choi}, Y.-Y. 2008, \bibinfo{title}{{Transformation of Morphology and Luminosity Classes of the SDSS Galaxies},} \apj, 674, 784, \dodoi{10.1086/524192}

\bibitem[{C. Park {et~al.}(2022)Park, Lee, Kim, Jeong, Pichon, Gibson, Snaith, Shin, Kim, Dubois, \& Few}]{Park_2022}
Park, C., Lee, J., Kim, J., {et~al.} 2022, \bibinfo{title}{Formation and Morphology of the First Galaxies in the Cosmic Morning,} The Astrophysical Journal, 937, 15, \dodoi{10.3847/1538-4357/ac85b5}

\bibitem[{ {Planck Collaboration} {et~al.}(2016){Planck Collaboration}, {Ade}, {Aghanim}, {Arnaud}, {Ashdown}, {Aumont}, {Baccigalupi}, {Banday}, {Barreiro}, {Bartlett}, {Bartolo}, {Battaner}, {Battye}, {Benabed}, {Beno{\^\i}t}, {Benoit-L{\'e}vy}, {Bernard}, {Bersanelli}, {Bielewicz}, {Bock}, {Bonaldi}, {Bonavera}, {Bond}, {Borrill}, {Bouchet}, {Boulanger}, {Bucher}, {Burigana}, {Butler}, {Calabrese}, {Cardoso}, {Catalano}, {Challinor}, {Chamballu}, {Chary}, {Chiang}, {Chluba}, {Christensen}, {Church}, {Clements}, {Colombi}, {Colombo}, {Combet}, {Coulais}, {Crill}, {Curto}, {Cuttaia}, {Danese}, {Davies}, {Davis}, {de Bernardis}, {de Rosa}, {de Zotti}, {Delabrouille}, {D{\'e}sert}, {Di Valentino}, {Dickinson}, {Diego}, {Dolag}, {Dole}, {Donzelli}, {Dor{\'e}}, {Douspis}, {Ducout}, {Dunkley}, {Dupac}, {Efstathiou}, {Elsner}, {En{\ss}lin}, {Eriksen}, {Farhang}, {Fergusson}, {Finelli}, {Forni}, {Frailis}, {Fraisse}, {Franceschi}, {Frejsel}, {Galeotta}, {Galli}, {Ganga}, {Gauthier}, {Gerbino}, {Ghosh}, {Giard},
  {Giraud-H{\'e}raud}, {Giusarma}, {Gjerl{\o}w}, {Gonz{\'a}lez-Nuevo}, {G{\'o}rski}, {Gratton}, {Gregorio}, {Gruppuso}, {Gudmundsson}, {Hamann}, {Hansen}, {Hanson}, {Harrison}, {Helou}, {Henrot-Versill{\'e}}, {Hern{\'a}ndez-Monteagudo}, {Herranz}, {Hildebrandt}, {Hivon}, {Hobson}, {Holmes}, {Hornstrup}, {Hovest}, {Huang}, {Huffenberger}, {Hurier}, {Jaffe}, {Jaffe}, {Jones}, {Juvela}, {Keih{\"a}nen}, {Keskitalo}, {Kisner}, {Kneissl}, {Knoche}, {Knox}, {Kunz}, {Kurki-Suonio}, {Lagache}, {L{\"a}hteenm{\"a}ki}, {Lamarre}, {Lasenby}, {Lattanzi}, {Lawrence}, {Leahy}, {Leonardi}, {Lesgourgues}, {Levrier}, {Lewis}, {Liguori}, {Lilje}, {Linden-V{\o}rnle}, {L{\'o}pez-Caniego}, {Lubin}, {Mac{\'\i}as-P{\'e}rez}, {Maggio}, {Maino}, {Mandolesi}, {Mangilli}, {Marchini}, {Maris}, {Martin}, {Martinelli}, {Mart{\'\i}nez-Gonz{\'a}lez}, {Masi}, {Matarrese}, {McGehee}, {Meinhold}, {Melchiorri}, {Melin}, {Mendes}, {Mennella}, {Migliaccio}, {Millea}, {Mitra}, {Miville-Desch{\^e}nes}, {Moneti}, {Montier}, {Morgante}, {Mortlock},
  {Moss}, {Munshi}, {Murphy}, {Naselsky}, {Nati}, {Natoli}, {Netterfield}, {N{\o}rgaard-Nielsen}, {Noviello}, {Novikov}, {Novikov}, {Oxborrow}, {Paci}, {Pagano}, {Pajot}, {Paladini}, {Paoletti}, {Partridge}, {Pasian}, {Patanchon}, {Pearson}, {Perdereau}, {Perotto}, {Perrotta}, {Pettorino}, {Piacentini}, {Piat}, {Pierpaoli}, {Pietrobon}, {Plaszczynski}, {Pointecouteau}, {Polenta}, {Popa}, {Pratt}, \& {Pr{\'e}zeau}}]{Planck2015}
{Planck Collaboration}, {Ade}, P.~A.~R., {Aghanim}, N., {et~al.} 2016, \bibinfo{title}{{Planck 2015 results. XIII. Cosmological parameters},} \aap, 594, A13, \dodoi{10.1051/0004-6361/201525830}

\bibitem[{P. {Popesso} {et~al.}(2024){Popesso}, {Biviano}, {Marini}, {Dolag}, {Vladutescu-Zopp}, {Csizi}, {Biffi}, {Lamer}, {Robothan}, {Bravo}, {Lovisari}, {Ettori}, {Angelinelli}, {Driver}, {Toptun}, {Dev}, {Mazengo}, {Merloni}, {Comparat}, {Ponti}, {Mroczkowski}, {Bulbul}, {Grandis}, \& {Bahar}}]{Popesso2024}
{Popesso}, P., {Biviano}, A., {Marini}, I., {et~al.} 2024, \bibinfo{title}{{The hot gas mass fraction in halos. From Milky Way-like groups to massive clusters},} arXiv e-prints, arXiv:2411.16555, \dodoi{10.48550/arXiv.2411.16555}

\bibitem[{M. {Raptis} {et~al.}(2025){Raptis}, {Rudie}, {Trainor}, {Rogers}, {Strom}, {Korhonen Cuestas}, {von Raesfeld}, {Lin}, {Ojodomo Abraham}, {Chapman}, {Steidel}, \& {Maseda}}]{Raptis2025}
{Raptis}, M., {Rudie}, G.~C., {Trainor}, R.~F., {et~al.} 2025, \bibinfo{title}{{CECILIA: The Mass-Metallicity Relation of Low-Mass Galaxies at Cosmic Noon},} arXiv e-prints, arXiv:2512.00162, \dodoi{10.48550/arXiv.2512.00162}

\bibitem[{A.~E. {Reines} \& M. {Volonteri}(2015){Reines} \& {Volonteri}}]{Reines2015}
{Reines}, A.~E., \& {Volonteri}, M. 2015, \bibinfo{title}{{Relations between Central Black Hole Mass and Total Galaxy Stellar Mass in the Local Universe},} \apj, 813, 82, \dodoi{10.1088/0004-637X/813/2/82}

\bibitem[{N. {Roy} {et~al.}(2018){Roy}, {Napolitano}, {La Barbera}, {Tortora}, {Getman}, {Radovich}, {Capaccioli}, {Brescia}, {Cavuoti}, {Longo}, {Raj}, {Puddu}, {Covone}, {Amaro}, {Vellucci}, {Grado}, {Kuijken}, {Verdoes Kleijn}, \& {Valentijn}}]{Roy2018}
{Roy}, N., {Napolitano}, N.~R., {La Barbera}, F., {et~al.} 2018, \bibinfo{title}{{Evolution of galaxy size-stellar mass relation from the Kilo-Degree Survey},} \mnras, 480, 1057, \dodoi{10.1093/mnras/sty1917}

\bibitem[{S. {Shen} {et~al.}(2003){Shen}, {Mo}, {White}, {Blanton}, {Kauffmann}, {Voges}, {Brinkmann}, \& {Csabai}}]{Shen2003}
{Shen}, S., {Mo}, H.~J., {White}, S. D.~M., {et~al.} 2003, \bibinfo{title}{{The size distribution of galaxies in the Sloan Digital Sky Survey},} \mnras, 343, 978, \dodoi{10.1046/j.1365-8711.2003.06740.x}

\bibitem[{M. {Shuntov} {et~al.}(2025){Shuntov}, {Ilbert}, {Toft}, {Arango-Toro}, {Akins}, {Casey}, {Franco}, {Harish}, {Kartaltepe}, {Koekemoer}, {McCracken}, {Paquereau}, {Laigle}, {Bethermin}, {Dubois}, {Drakos}, {Faisst}, {Gozaliasl}, {Gillman}, {Hayward}, {Hirschmann}, {Huertas-Company}, {Jespersen}, {Jin}, {Kokorev}, {Lambrides}, {Le Borgne}, {Liu}, {Magdis}, {Massey}, {McPartland}, {Mercier}, {McCleary}, {McKinney}, {Oesch}, {Renzini}, {Rhodes}, {Rich}, {Robertson}, {Sanders}, {Trebitsch}, {Tresse}, {Valentino}, {Vijayan}, {Weaver}, {Weibel}, {Wilkins}, \& {Yang}}]{Shuntov2025}
{Shuntov}, M., {Ilbert}, O., {Toft}, S., {et~al.} 2025, \bibinfo{title}{{COSMOS-Web: Stellar mass assembly in relation to dark matter halos across 0.2 < z < 12 of cosmic history},} \aap, 695, A20, \dodoi{10.1051/0004-6361/202452570}

\bibitem[{R.~S. {Sutherland} \& M.~A. {Dopita}(1993){Sutherland} \& {Dopita}}]{Sutherland1993}
{Sutherland}, R.~S., \& {Dopita}, M.~A. 1993, \bibinfo{title}{{Cooling Functions for Low-Density Astrophysical Plasmas},} \apjs, 88, 253, \dodoi{10.1086/191823}

\bibitem[{R. {Teyssier}(2002){Teyssier}}]{Teyssier2002}
{Teyssier}, R. 2002, \bibinfo{title}{{Cosmological hydrodynamics with adaptive mesh refinement. A new high resolution code called RAMSES},} \aap, 385, 337, \dodoi{10.1051/0004-6361:20011817}

\bibitem[{C. {Tortora} {et~al.}(2025){Tortora}, {Busillo}, {Napolitano}, {Koopmans}, {Covone}, {Genel}, {Villaescusa-Navarro}, \& {Silvestrini}}]{Tortora2025}
{Tortora}, C., {Busillo}, V., {Napolitano}, N.~R., {et~al.} 2025, \bibinfo{title}{{CASCO: Cosmological and AStrophysical parameters from Cosmological simulations and Observations III. The physics behind the emergence of the golden mass scale},} arXiv e-prints, arXiv:2502.13589, \dodoi{10.48550/arXiv.2502.13589}

\bibitem[{C.~A. {Tremonti} {et~al.}(2004){Tremonti}, {Heckman}, {Kauffmann}, {Brinchmann}, {Charlot}, {White}, {Seibert}, {Peng}, {Schlegel}, {Uomoto}, {Fukugita}, \& {Brinkmann}}]{Tremonti2004}
{Tremonti}, C.~A., {Heckman}, T.~M., {Kauffmann}, G., {et~al.} 2004, \bibinfo{title}{{The Origin of the Mass-Metallicity Relation: Insights from 53,000 Star-forming Galaxies in the Sloan Digital Sky Survey},} \apj, 613, 898, \dodoi{10.1086/423264}

\bibitem[{T. {Tsukui} {et~al.}(2024){Tsukui}, {Wisnioski}, {Bland-Hawthorn}, \& {Freeman}}]{Tsukui2024}
{Tsukui}, T., {Wisnioski}, E., {Bland-Hawthorn}, J., \& {Freeman}, K. 2024, \bibinfo{title}{{The emergence of galactic thin and thick discs across cosmic history},} arXiv e-prints, arXiv:2409.15909, \dodoi{10.48550/arXiv.2409.15909}

\bibitem[{A. {van der Wel}(2008){van der Wel}}]{wel2008}
{van der Wel}, A. 2008, \bibinfo{title}{{The Dependence of Galaxy Morphology and Structure on Environment and Stellar Mass},} \apjl, 675, L13, \dodoi{10.1086/529432}

\bibitem[{R.~H. Wechsler \& J.~L. Tinker(2018)Wechsler \& Tinker}]{Wechsler_2018}
Wechsler, R.~H., \& Tinker, J.~L. 2018, \bibinfo{title}{The Connection Between Galaxies and Their Dark Matter Halos,} Annual Review of Astronomy and Astrophysics, 56, 435–487, \dodoi{10.1146/annurev-astro-081817-051756}

\bibitem[{A.~R. {Wetzel} {et~al.}(2013){Wetzel}, {Tinker}, {Conroy}, \& {van den Bosch}}]{Wetzel2013}
{Wetzel}, A.~R., {Tinker}, J.~L., {Conroy}, C., \& {van den Bosch}, F.~C. 2013, \bibinfo{title}{{Galaxy evolution in groups and clusters: satellite star formation histories and quenching time-scales in a hierarchical Universe},} \mnras, 432, 336, \dodoi{10.1093/mnras/stt469}

\bibitem[{Y. Yoon {et~al.}(2021)Yoon, Park, Chung, \& Zhang}]{Yoon_2021}
Yoon, Y., Park, C., Chung, H., \& Zhang, K. 2021, \bibinfo{title}{Rotation Curves of Galaxies and Their Dependence on Morphology and Stellar Mass,} The Astrophysical Journal, 922, 249, \dodoi{10.3847/1538-4357/ac2302}

\end{thebibliography}
\bibliographystyle{aasjournal}



\end{document}